\documentclass[conference]{IEEEtran/IEEEtran} %

\usepackage{microtype} %
\usepackage{balance} %
\usepackage{algorithmic}
\usepackage[table]{xcolor} %
\usepackage{xspace} %
\usepackage{subcaption} %
\usepackage{adjustbox} %
\usepackage{listings}
\definecolor{codegreen}{rgb}{0,0.6,0}
\definecolor{codegray}{rgb}{0.5,0.5,0.5}
\definecolor{codepurple}{rgb}{0.58,0,0.82}
\definecolor{backcolour}{rgb}{0.95,0.95,0.92}

\lstdefinestyle{mystyle}{
	backgroundcolor=\color{white},
	keywordstyle=\color{codegreen},
	numberstyle=\tiny\color{codegray},
	stringstyle=\color{codepurple},
	basicstyle=\ttfamily,
	breakatwhitespace=false,
	breaklines=true,
	captionpos=b,
	keepspaces=true,
	numbers=left,
	numbersep=5pt,
	showspaces=false,
	showstringspaces=false,
	showtabs=false,
	tabsize=2,
    morekeywords=[1]{
    },
}

\lstdefinestyle{customMLIR}{
    inputencoding=utf8,
    tabsize=2,
    rulecolor=,
    upquote=true,
    columns=fixed,
    linewidth=\columnwidth,
    showstringspaces=false,
    extendedchars=true,
    breaklines=true,
    showtabs=false,
    showspaces=false,
    showstringspaces=false,
    basicstyle=\scriptsize\ttfamily,
    identifierstyle=\scriptsize\ttfamily,
    keywordstyle=\scriptsize\ttfamily\color[rgb]{0,0,1},
    commentstyle=\scriptsize\ttfamily\color[rgb]{0.133,0.545,0.133},
    stringstyle=\scriptsize\ttfamily\color[rgb]{0.627,0.126,0.941},
}

\makeatletter

\lstdefinelanguage{mlir}{
  morecomment = [l]{//},
  morestring=[b]", 
  sensitive = true,
  classoffset=0,
  classoffset=1, keywordstyle=\color{purple},
  morekeywords={
    soda.launch_func,
    soda.module,
    soda.return,
    func,
    define, declare, global, constant,
    internal, external, private,
    linkonce, linkonce_odr, weak, weak_odr, appending,
    common, extern_weak,
    thread_local, dllimport, dllexport,
    hidden, protected, default,
    except, deplibs,
    volatile, fastcc, coldcc, cc, ccc,
    x86_stdcallcc, x86_fastcallcc,
    ptx_kernel, ptx_device,
    signext, zeroext, inreg, sret, nounwind, noreturn,
    nocapture, byval, nest, readnone, readonly, noalias, uwtable,
    inlinehint, noinline, alwaysinline, optsize, ssp, sspreq,
    noredzone, noimplicitfloat, naked, alignstack,
    module, asm, align, tail, to,
    addrspace, section, alias, sideeffect, c, gc,
    target, datalayout, triple,
    blockaddress,
    return,
    step,
    func.func, scf.for, 
    linalg.matmul, linalg.generic,
    memref.load, memref.store, memref.subview,
    addi, muli, addf, mulf
  },
  classoffset=2, keywordstyle=\color{gray},
  morekeywords={
    memref, index, f32,
    !iarr_t, !varr_t,
    !AH_t, !AHW_t,
    !mr4x4_0, !mr4x4_1
  },
  alsoletter={\%, ., \!},
  keywordsprefix={\%},
}
\makeatother %
\definecolor{delim}{RGB}{20,105,176}
\definecolor{numb}{RGB}{106, 109, 32}
\definecolor{string}{rgb}{0.64,0.08,0.08}

\lstdefinelanguage{json}{
    tabsize=2,
    numbers=left,
    numberstyle=\small,
    rulecolor=\color{black},
    showspaces=false,
    showtabs=false,
    breaklines=true,
    postbreak=\raisebox{0ex}[0ex][0ex]{\ensuremath{\color{gray}\hookrightarrow\space}},
    breakatwhitespace=true,
    basicstyle=\scriptsize\ttfamily,
    upquote=true,
    morestring=[b]",
    stringstyle=\scriptsize\ttfamily\color[rgb]{0.627,0.126,0.941},
    numberstyle=\tiny,
    texcl=false
    literate=
     *{0}{{{\color{numb}0}}}{1}
      {1}{{{\color{numb}1}}}{1}
      {2}{{{\color{numb}2}}}{1}
      {3}{{{\color{numb}3}}}{1}
      {4}{{{\color{numb}4}}}{1}
      {5}{{{\color{numb}5}}}{1}
      {6}{{{\color{numb}6}}}{1}
      {7}{{{\color{numb}7}}}{1}
      {8}{{{\color{numb}8}}}{1}
      {9}{{{\color{numb}9}}}{1}
      {\{}{{{\color{delim}{\{}}}}{1}
      {\}}{{{\color{delim}{\}}}}}{1}
      {[}{{{\color{delim}{[}}}}{1}
      {]}{{{\color{delim}{]}}}}{1},
  } %
\lstdefinestyle{customSyntax}{
    inputencoding=utf8,
    tabsize=2,
    rulecolor=,
    upquote=true,
    columns=fixed,
    linewidth=\columnwidth,
    showstringspaces=false,
    extendedchars=true,
    breaklines=true,
    showtabs=false,
    showspaces=false,
    showstringspaces=false,
    basicstyle=\scriptsize\ttfamily,
    identifierstyle=\scriptsize\ttfamily,
    keywordstyle=\scriptsize\ttfamily\color[rgb]{0,0,1},
    commentstyle=\scriptsize\ttfamily\color[rgb]{0.133,0.545,0.133},
    stringstyle=\scriptsize\ttfamily\color[rgb]{0.627,0.126,0.941},
}

\makeatletter

\lstdefinelanguage{syntax}{
  morecomment = [l]{//},
  morestring=[b]", 
  sensitive = true,
  classoffset=0,
  classoffset=1, keywordstyle=\color{purple},
  morekeywords={
     opcode_dict,
     opcode_entry,
     opcode_expr,
     opcode_flow_entry,
     bare_id,
     string_literal,
     opcode_list,
     integer_literal,
     flow_expr,
  },
  classoffset=2, keywordstyle=\color{gray},
  morekeywords={
    memref, index, f32,
    !iarr_t, !varr_t,
    !AH_t, !AHW_t,
    !mr4x4_0, !mr4x4_1
  },
  alsoletter={\%, ., \!},
  keywordsprefix={\%},
}
\makeatother %
\usepackage{multirow} %
\usepackage{blindtext} 
\usepackage[bookmarks=false]{hyperref}%

\usepackage{tikz} %
\newcommand*\circled[1][]{\tikz[baseline=(char.base)]{
            \node[shape=circle,draw,inner sep=1.5pt] (char) {\footnotesize#1};}}
            
\usepackage[normalem]{ulem}
\useunder{\uline}{\ul}{}

\definecolor{cadmiumgreen}{rgb}{0.0, 0.42, 0.24}
\definecolor{cardinal}{rgb}{0.77, 0.12, 0.23}
\definecolor{lightgray}{gray}{0.9}
\definecolor{lightblue}{rgb}{0.93,0.95,1.0}
\definecolor{royalpurple}{rgb}{0.47,0.32,0.66}
\definecolor{richlilac}{rgb}{0.71, 0.4, 0.82}
\definecolor{veronica}{rgb}{0.63, 0.36, 0.94}

\newcommand{\mlir}{MLIR\xspace}

\newcommand{\linalg}{\lstinline{linalg}\xspace}

\newcommand{\scf}{\lstinline{scf}\xspace}

\newcommand{\added}[1]{{\color{black}#1}}
\newcommand{\newtext}[1]{{\color{black}#1}}
\newcommand{\newt}[1]{{\color{black}#1}} %

\begin{document}

\title{AXI4MLIR: User-Driven Automatic Host Code Generation for Custom AXI-Based~Accelerators}

\author{\IEEEauthorblockN{Nicolas Bohm Agostini\IEEEauthorrefmark{2}\IEEEauthorrefmark{5},
Jude Haris\IEEEauthorrefmark{3}, \\
Perry Gibson\IEEEauthorrefmark{3},
Malith Jayaweera\IEEEauthorrefmark{2},
Norm Rubin\IEEEauthorrefmark{2}, \\
Antonino Tumeo\IEEEauthorrefmark{5},
José L. Abellán\IEEEauthorrefmark{4},
José Cano\IEEEauthorrefmark{3},
David Kaeli\IEEEauthorrefmark{2}
}
\IEEEauthorblockA{
\IEEEauthorrefmark{2}\textit{Northeastern University},
Boston, MA, USA 
\IEEEauthorrefmark{3}\textit{University of Glasgow},
Glasgow, Scotland, UK \\
\IEEEauthorrefmark{4}\textit{University of Murcia},
Murcia, Spain
\IEEEauthorrefmark{5}\textit{Pacific Northwest National Laboratory},
Richland, WA, USA
}}

\IEEEoverridecommandlockouts
\IEEEpubid{\begin{minipage}{\textwidth}\ \\[12pt]
Nicolas Bohm Agostini and Jude Haris are co-first authors \\
\end{minipage}} 

\maketitle %

\begin{abstract}

This paper addresses the need for automatic and efficient generation of host driver code for arbitrary custom AXI-based accelerators targeting linear algebra algorithms, an important workload in various applications, including machine learning and scientific computing. While existing tools have focused on automating accelerator prototyping, little attention has been paid to the host-accelerator interaction. This paper introduces AXI4MLIR, an extension of the MLIR compiler framework designed to facilitate the automated generation of host-accelerator driver code. With new MLIR attributes and transformations, AXI4MLIR empowers users to specify accelerator features (including their instructions) and communication patterns and exploit the host memory hierarchy. We demonstrate AXI4MLIR's versatility across different types of accelerators and problems, showcasing significant CPU cache reference reductions (up to 56\%) and up to a 1.65$\times$ speedup compared to manually optimized driver code implementations. AXI4MLIR implementation is open-source and available at: \url{https://github.com/AXI4MLIR/axi4mlir}.

\end{abstract}


\section{Introduction}

Given the diminishing performance gains provided by today's general-purpose computing~\cite{henessy2018goldenage}, there has been renewed interest in exploring custom hardware accelerators. Accelerators can support architecture-level optimizations that can increase the performance and efficiency of key applications~\cite{shabani2023hpca,kim2023hpca,zhao2022isca,hsia2023asplos,Zheng2022AMOSEA,munoz2023asplos}. One important class of applications that can benefit from accelerators is tensor algebra processing, which is widely used in the domains of machine learning, scientific computing, and data analytics~\cite{abolhasani2023sdl,Rao2018sdc,jumper2021alphafold}. Tensor operations tend to be computationally intensive and require high memory bandwidth, making them suitable for specialized hardware implementations. Automated tools have been proposed~\cite{Zhang2020dnnexplorer,pengfei2020autoDNNchip,ye2020HybridDNN,kwon2020maestro} to help explore new classes of custom domain-specific accelerators targeting tensor computations, and are currently the best path available to obtain performance gains in scientific workloads and machine learning applications.

However, designing and fully exploiting custom hardware accelerators for tensor operations is not a trivial task~\cite{gibsonDLAS2023}. When co-designing these devices, we need to generate efficient architectures, and we must optimize the communication between the host CPU and the accelerator. In particular, the host-accelerator interaction involves several aspects, including data transfers, synchronization, and the accelerator's control flow. These aspects depend on the characteristics of the host CPU microarchitecture, the host-accelerator interface, the accelerator design, and the application code. 
Manually rewriting the host driver code for each accelerator and application scenario can be very tedious and error-prone. Furthermore, most of the prior work proposing new accelerators~\cite{chen2019eyeriss,tvm2020vta,loncar2020hls4ml,Skalicky2018Hot,agostini2020tflitesoc} only considers a simple offload model or assumes that the required data is already placed in the accelerator's internal buffers, falling short in providing insights into how host-to-accelerator transfers should be performed or generated. 
Additionally, complex accelerators, exemplified by Google's TPUs and Nvidia's GPUs, benefit from large teams that can collaboratively engineer dedicated compilers to address some of these issues. 
However, smaller development teams may lack expertise or available time resources to invest in compilers.
Consequently, custom accelerator designers typically implement driver code and instruction streams manually to validate and deploy their designs for a subset of synthetic workloads.

\begin{figure}[t]
\centering
\includegraphics[width=.9\linewidth]{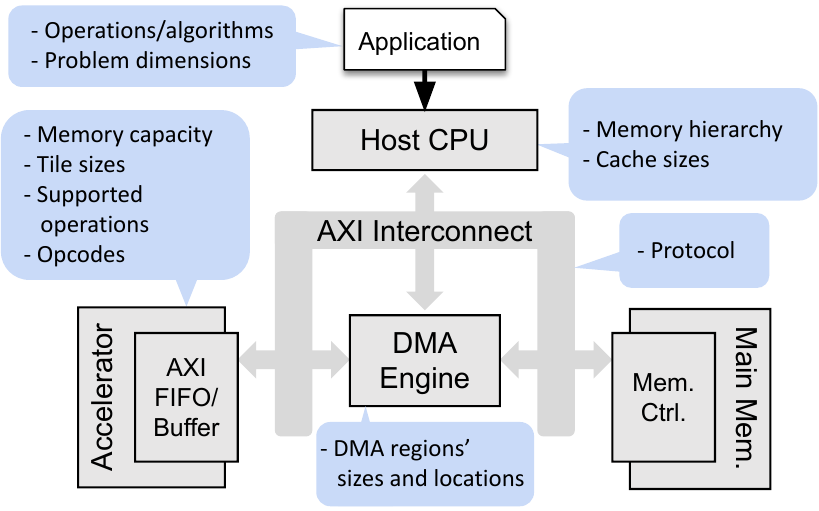}
\caption{Typical host-accelerator system design, highlighting (blue color) relevant parameters that should be considered for efficient generation of host-accelerator communication code.}
\label{fig:sys_design}
\end{figure}

To implement or generate efficient host-to-accelerator communication, we argue that it is necessary to consider all major features of a System-on-Chip (SoC). Figure~\ref{fig:sys_design} highlights a typical system using an AXI~\cite{arm2003axi} interconnection between the CPU and a custom accelerator, which is a common choice in many designs~\cite{liu2018trets}. To drive the accelerator effectively, the host-code implementation should exploit features regarding the CPU, the interconnect, and the accelerator (see Figure~\ref{fig:sys_design}).

To effectively consider each of the key system features described in Figure~\ref{fig:sys_design} while also delivering efficient and automated CPU-accelerator driver code generation, we propose \textbf{AXI4MLIR}, an extension to the MLIR compiler framework~\cite{Lattner2021mlir} that enables efficient and automated CPU-accelerator driver code generation for accelerators targeting linear algebra applications. AXI4MLIR takes a high-level application description in the MLIR’s linear algebra (\lstinline{linalg}) abstraction~\cite{mlir2020linalg} as input and introduces custom MLIR attributes to describe the target accelerator capabilities. These attributes provide accelerator-specific information to custom transformation passes that can effectively specialize and generate accelerator-aware host driver code.
Our extensions facilitate hardware-software co-design by allowing developers to automatically generate driver code with varying configurations, more easily explore their design space, and use the designed accelerator in applications that can be compiled with the MLIR framework.
The contributions of this work include the following:\looseness=-1 

\begin{itemize}
    
    \item New MLIR attributes that provide a standardized and extensible approach to represent accelerators that can implement a range of linear algebra algorithms supported by the MLIR \textit{linalg} abstraction.

    \item Automated generation of efficient driver code for custom accelerators leveraging AXI-based interfaces in host-to-accelerator communication. 
    
    \item The ability to describe and explore accelerator-specific tiling and dataflow strategies for the target linear algebra operation, which can improve computation efficiency within the accelerator and reduce data movement overheads between the accelerator and CPU.
    
    \item An analysis of our compiler optimizations on a suite of benchmarks representing key linear algebra applications, demonstrating the effectiveness of our approach in achieving significant performance gains (up to 1.65$\times$ speedup and 56\% fewer cache references) when compared to optimized manual driver code implementations.

\end{itemize}

While leveraging the new attributes of AXI4MLIR, our user-directed host code generation is entirely automated by the compiler. This provides a significant advantage in terms of productivity and maintainability.

\section{Background}\label{sec:background}

\subsection{MLIR}\label{sec:sub-mlir}

\mlir is a compiler infrastructure framework that facilitates the creation of domain-specific compilers by providing code generators, translators, optimizers, and the infrastructure to define subsets of operations that expose well-defined language abstractions~\cite{Lattner2021mlir,Le2020onnxmlir}. 
Notably, \mlir offers support for compilation from various frontends into its infrastructure, encompassing frameworks such as TensorFlow, PyTorch, and ONNX, as well as languages like Fortran, C, and Mojo.
In \mlir, a group of operations modeling an abstraction is called a \textit{dialect}. Dialects are self-contained intermediate representations (IRs) and follow the language rules of \mlir's meta-IR, enabling the framework to have multiple dialects coexisting in the same \mlir file. This approach promotes the reuse of already defined abstractions and associated tools, enabling intra- and inter-dialect transformations.

\begin{figure}[t]
\centering
\begin{adjustbox}{minipage=\linewidth,scale=0.9}

\begin{subfigure}[t]{1\textwidth}
\centering
\input{lsts/matmul-mlir-linalg-generic}
\caption{Linalg Abstraction with generic operation.}
\label{fig:lst-mlir-matmul-generic}
\end{subfigure}

\begin{subfigure}[t]{1\textwidth}
\centering
\input{lsts/matmul-mlir-scf}
\caption{\added{Structured Control Flow (SCF)} Abstraction with tiling.}\label{fig:lst-mlir-matmul-scf}
\end{subfigure}
\end{adjustbox}
\caption{\newt{MLIR representations of a Matrix-Matrix Multiplication Operation in different abstractions.\textsuperscript{1}}}
\label{fig:mlir-matmul-kernel}
\end{figure}

In support of the underlying algorithms and kernels used by many machine learning frameworks (e.g., TensorFlow and PyTorch), \mlir offers a linear algebra dialect called \linalg that exposes (named) operations such as convolutions, matrix multiplications, and others. Operations expressed in higher-level dialects can target \linalg operations and leverage all subsequent transformations supported by \linalg and lower-level abstractions. Figure~\ref{fig:mlir-matmul-kernel} presents an \mlir matrix-multiplication (MatMul) implementation in different abstractions.\footnote{\newt{We intentionally omit some MLIR code, such as constant declarations in the form of \texttt{\%cX=arith.constant X:i32}, for the sake of brevity.}} 
The operation is initially represented using a \lstinline{linalg.matmul} and subsequently undergoes conversion, transformation, and lowering by the compiler.
In Figure~\ref{fig:lst-mlir-matmul-generic}-L11 and L17, the \lstinline{linalg.matmul} is converted into a \lstinline{linalg.generic}. The \lstinline{linalg.generic} is a core \mlir operation that can represent most of the \lstinline{linalg} \textit{named ops}, by careful selection of its \textit{operation trait}\footnote{See \textit{linalg.generic} in https://mlir.llvm.org/docs/Dialects/Linalg} - \lstinline{indexing_maps} (L2), \lstinline{iterator_types} (L7) -, and kernel (L24 to L27). Finally, the generic operation can be converted into a tiled (4$\times$4$\times$4) implementation of the MatMul (Figure~\ref{fig:lst-mlir-matmul-scf}) using the structured control flow (\scf) dialect. When supporting an accelerator that can process a MatMul$_{4x4x4}$ operation~\footnote{A 2D MatMul operation is MatMul$_{MxNxK}$: C(M,N) = A(M,K) x B(K,N)}, the code in Figure~\ref{fig:lst-mlir-matmul-scf}-L11 to L19, has to be replaced by the runtime library calls that drive the accelerator.

\subsubsection{MLIR Memory References}\label{sec:mlir-memrefs}

Within MLIR, memory buffers exist as N-dimensional (rank=N) memory references, or \mbox{\lstinline{memref}s}. Our proposed AXI4MLIR DMA runtime library, presented in Section~\ref{sec:dma-rt-lib}, supports bidirectional data movements between \lstinline{memrefs} and memory-mapped buffers (raw pointers), while respecting strides, sizes, and dimensions. Accessing the elements of an MLIR \lstinline{memref} requires accessing the values in the equivalent C struct of Figure~\ref{fig:lst-memref-cstruct}. Specializing the code for specific sizes and strides is an important proposed optimization to leverage spatial locality and minimize control-flow instructions, as we will observe in Section~\ref{sec:results}.

\begin{figure}[htbp]
\centering
\begin{adjustbox}{minipage=\linewidth,scale=0.75}
\begin{lstlisting}[language=C,basicstyle=\small\ttfamily]
typedef struct {
  float *allocated; // For deallocation
  float *aligned;   // Base address
  size_t offset;    // Offset in # of elements
  size_t size[N];   // One size per dim
  size_t stride[N]; // One stride per dim
}
\end{lstlisting}
\end{adjustbox}
\caption{The underlying data structure of a rank==N MLIR \lstinline{memref} buffer.}\label{fig:lst-memref-cstruct}
\vspace{-10pt}
\end{figure}

\subsection{AXI Interface}

Efficiently using the interconnect between the CPU and the accelerator can significantly impact the overall system performance. As part of our framework, we consider a widely adopted bus interface in digital electronics design deployed on SoC and Field-Programmable Gate Array (FPGA) designs, namely Advanced eXtensible Interface (AXI)~\cite{arm2003axi}. AXI provides a flexible and scalable solution for integrating custom accelerators into a system.

The AXI interface provides a simple mechanism to enable data transfers between the CPU cores and other devices.
Using AXI, the AXI-Stream (AXI-S) interface allows the developer to quickly transfer~\cite{axiguide} a variable-size burst of data to and from the accelerator in a FIFO-like manner, enabling the accelerator to consume/store the data as needed, in a streaming manner. Within SoCs, the CPU host code controls either a single or multiple Direct Memory Access (DMA) engines (see Figure~\ref{fig:sys_design}). These engines are responsible for initiating and handling data movement requests between the main memory and the accelerator. Additionally, the data regions in the main memory need to be accessible to the accelerator via the \mbox{AXI-S} interface. Therefore, the host code needs to allocate input and output memory buffers using the \textit{mmap} function, which guarantees that only the current process has access to the specific regions of memory. The host code is also required to prepare/pack the input data into the data format that the accelerator requires (e.g., row-major, interleaved data elements, etc.). Our approach within AXI4MLIR is to \textit{use MLIR - the compiler - to generate the host code to interface with the accelerator}, while taking advantage of the full capabilities of the target accelerator.

\section{AXI4MLIR}\label{sec:sub-axi4mlir}

\begin{figure}[t]
\centering
\includegraphics[width=1\linewidth]{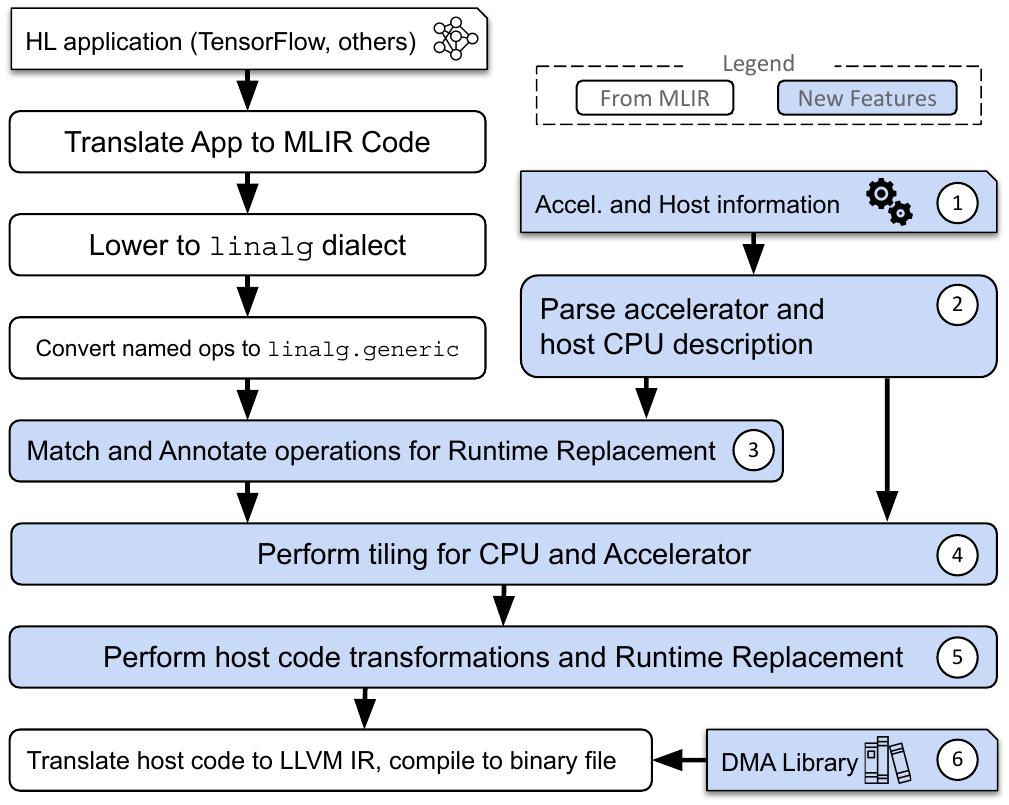}
\caption{AXI4MLIR Compiler Flow. The numbered elements are the contributions of this work.}\label{fig:compiler-flow}
\end{figure}

To support efficient host code generation for AXI-based custom accelerators, we extended the MLIR framework with the added capabilities presented in Figure~\ref{fig:compiler-flow}. After the custom accelerator is designed and the host CPU system is selected, the user creates a configuration file with the host CPU system details (e.g., number and size of the caches), and with a high-level description of the accelerator capabilities (i.e., supported operations and dimensions), the available opcodes, and possible opcode flows \circled[1]. This information is parsed \circled[2] by the compiler, and used to find \circled[3] suitable \lstinline{linalg.generic} operations with the desired operation traits (algorithm implemented, previously shown in Figure~\ref{fig:lst-mlir-matmul-generic}-L1 to L9), that can be executed on the accelerator. These operations will require host-accelerator driver code generation. Subsequently, with user-provided information on the total size of the CPU caches, the compiler transforms the code to efficiently exploit the CPU memory hierarchy and the accelerator size \circled[4], performing the appropriate set of tiling transformations to leverage temporal locality in the CPU caches and to map the problem on the accelerator. In the final step, the compiler generates the runtime calls \circled[5] that leverage the accelerator features based on user-directed dataflow description (e.g., avoiding redundant host-accelerator data transfers when the algorithm and accelerator functionality allows).

The following sections discuss the class of supported accelerators and the key features of our AXI4MLIR DMA library. We provide details on how to describe new accelerators, introducing \lstinline{linalg.generic} trait extensions, a new MLIR dialect that provides support for runtime call replacement of opcodes and data transfers, and some key optimizations that can be performed (depending on the available features of the host system and the custom accelerator).

\subsection{The Custom AXI DMA Library}\label{sec:dma-rt-lib}

The AXI4MLIR \emph{DMA library}~\circled[6] (Figure~\ref{fig:compiler-flow}) exposes low-level DMA calls working at privileged level to enable data movement between the main memory and the accelerator. 
We designed this library to be lightweight (55 bytes in size for our target ARM SoC), so that it can be deployed on both resource-constrained and non-constrained systems. It can also be executed by bare-metal systems. During the compilation process, the AXI runtime issues calls to initialize the DMA engine(s) before entering the computation kernel of the workload. First, a library call initializes the DMA engine, mapping memory for the input and output buffers which act as temporary staging buffers between the CPU and the accelerator.  

After DMA initialization, the accelerator is accessible via AXI-based data transfers. Any data that needs to be transferred to the accelerator during workload execution is first copied to a DMA input buffer. This staging copy acts as a packing optimization (similar to \cite{rohwedder2023gpat}), contributing to an increased cache-hit ratio during communication. Then, the AXI ``send'' function call requests the DMA engine to start the data transfer and waits for it to finish. Note that the data that is sent to the accelerator can be either accelerator instructions or raw input data that needs to be processed. Similarly, 
AXI4MLIR generates ``recv'' function calls to wait for computation completion and to obtain output data from the DMA output buffer.

\newtext{
In Section~\ref{sec:mlir-compiler-extension}, Figure~\ref{fig:accel-diagram} presents the lowering of different high-level operations into our DMA library calls. 
\lstinline{copy_to_dma_region(...)} implements data movement from a \lstinline{memref} to the DMA-accessible memory region intended for transmission to the accelerator. The \lstinline{offset} argument allows for efficient batching of different data transfers after computing the total length and executing a single ``send'' operation. Appropriate offset values prevent overwriting existing data in the DMA region.
\lstinline{dma_start_send(...)} instructs the DMA engine to transmit a \lstinline{size} of $X$ bytes to the connected accelerator, commencing from a specified \lstinline{offset} within the DMA space allocated. \lstinline{dma_wait_send_completion(...)} instructs the CPU to wait for the DMA's signal informing the transaction's completion.
When receiving data from the accelerator, we first have to wait for the data to be placed in the DMA-accessible memory so it can be  copied back into a \lstinline{memref}.}

\subsection{Supported Accelerators} 

\newtext{In matrix-multiplication and similar algorithms, the term \textit{stationary} refers to a slice of data that can be reused across many iterations of an algorithm's computation. A \textit{stationary} strategy attempts to maximize data reuse and minimize data movement, which can greatly benefit accelerators that require efficient memory accesses. We want to enable the programmer to easily control accelerators that support \textit{stationary} flows.}

Next, we discuss the types of accelerators that AXI4MLIR can support. Then we propose a standardized approach to concisely define the class of supported accelerators in a configuration file. 
Finally, we show how the AXI4MLIR parser is able to take user-defined configurations, extract essential attributes of the target accelerator, and populate a trait specification to guide our MLIR compiler transformations.

\subsubsection{Accelerator Designs}

The AXI4MLIR compiler transformations support linear algebra kernels implemented as accelerators using the AXI interconnect. In addition, the \mbox{AXI-S} data transfers within AXI4MLIR facilitate support for accelerators that use a micro-ISA (Instruction Set Architecture) with opcodes, which consist of instructions that the host-CPU sends to the accelerator.
Generally, the following three actions are used to categorize the actions within an instruction: \textit{send}, \textit{compute}, and \textit{receive}.
Any accelerator's instructions that require external communication (i.e., data transfers or activation/reset/configuration of accelerator compute modules) can be completed by issuing a combination of these three actions. In addition, each action can have additional meta-data (e.g., opcode literal, data, length, dimensions, and indexes), which is used to guide compiler transformations during accelerator host code generation. Further, specific traits of the accelerator - such as internal buffer space (or accelerator tile sizes), and data types - are supported and must be defined within the accelerator configuration file.

\subsubsection{Accelerator Configuration File}

\begin{figure}[hb]
\centering
\begin{adjustbox}{minipage=\linewidth,scale=0.95}
\centering
\input{lsts/acc_cpu_config}
\caption{Accelerator and CPU configuration file.}\label{fig:acc_cpu_config}
\end{adjustbox}
\end{figure}

Once an AXI-based accelerator is fully designed, the accelerator developer can quickly integrate it with our AXI4MLIR compiler transformations by providing \emph{Accelerator and Host information}~\circled[1] (Figure~\ref{fig:compiler-flow}) through a configuration file for the new accelerator and the target host system. Figure~\ref{fig:acc_cpu_config} shows a sample configuration file defined in the standard JSON format. For the accelerator, the developer must specify the accelerator's architectural features, e.g., supported tile sizes, data type, and input and output data with related dimensions. Additionally, the developer should describe any micro-ISA that the accelerator can execute. The developer should define ``opcode IDs'', captured by the ``opcode\_map string'', which are comprised of actions to describe the memory operations and related data transfers. Finally the developer should define the possible ``opcode flow IDs'' and select the desired flow for the particular operation. The configuration file does not capture the internal behavior of the accelerator, which has been the focus of other works~\cite{pengfei2020autoDNNchip,chen2019eyeriss}; instead, we seek to optimize the communication with the accelerator. Thus the configuration file contains information about the I/O interface for sending data and instructions to the accelerator. Similar to the accelerator information, the CPU information, shown in Figure~\ref{fig:acc_cpu_config}-L1 to L2, needs to contain basic architectural details such as the number and size of caches.

\subsubsection{Configuration Parsing}

The parser implemented in \circled[2] (Figure~\ref{fig:compiler-flow}) is responsible for providing the information from the configuration file to the MLIR IR and the AXI4MLIR transformation passes. To this end, the \textit{kernel} and \textit{cache information}, paired with a simple heuristic that identifies the dimensions of the target MLIR operation, are used to schedule tiling transformations (Figure~\ref{fig:compiler-flow}~-~\circled[4]) that leverage the CPU memory hierarchy sizes and increase temporal locality of the memory accesses. 
Additionally, the parser validates the \lstinline{opcode_map} and the user selected \lstinline{opcode_flow}, which are then translated into new MLIR attributes to the target \lstinline{linalg.generic} operation trait. Their syntax and functionality are described in Section~\ref{sec:mlir-compiler-extension}.

\added{

\subsubsection{Supported Systems}

Our work is focused on SoCs with accelerators connected to ARM CPUs via an AXI-S interconnect. AXI4MLIR seamlessly integrates with a diverse set of Xilinx platforms, though we also anticipate similar applicability to other FPGA-SoC devices. Changing the cross-compiler would allow support for other processors. Adapting our DMA library implementation to other standards would be required to support other types of interconnects. AXI4MLIR currently supports AXI-Stream accelerators, which do not communicate via direct memory requests. Thus, AXI4MLIR does not require support for host-accelerator coherence protocols, since the host manages the DMA engine transfers.
}

\subsection{MLIR extensions and optimizations}\label{sec:mlir-compiler-extension}

\begin{figure}[t]
\centering
\begin{adjustbox}{minipage=\columnwidth,scale=0.9}\begin{subfigure}[t]{1\textwidth}
\centering
\input{lsts/additional-traits}
\caption{New Attributes for Accelerator Description.}\label{fig:new-trait}
\end{subfigure}

\begin{subfigure}[t]{1\textwidth}
\centering
\input{lsts/matmul-with-accelops}
\caption{\added{\newt{IR to drive the MatMul accelerator with an A-stationary flow.}}}\label{fig:matmul-with-accel}
\end{subfigure}\end{adjustbox}
\caption{Information added to the linalg.generic traits to capture accelerator behavior in MLIR \added{and IR with accel operations.}}
\label{fig:mlir-matmul-kernel-traits}
\end{figure}

To implement \emph{match and annotate operations for runtime replacement}~\circled[3] (Figure~\ref{fig:compiler-flow}), and to offload the computation onto the accelerator, we implemented passes to identify the target algorithms supported by the accelerator and extended the \lstinline{linalg.generic} operation \lstinline{trait} with additional information, as shown in Figure~\ref{fig:new-trait}. In particular, we introduced two new types of attributes to MLIR, \lstinline{opcode_map} and \lstinline{opcode_flow}, which follow the syntax described in Figure~\ref{fig:opcode_map_syntax} and Figure~\ref{fig:opcode_flow_syntax}, respectively.
We elaborate more on each attribute in the operation \lstinline{trait} below.

\noindent\textbf{Extensions to linalg.generic traits:} %

\noindent \lstinline{- dma_init_config}: defines the parameter values used to configure a DMA engine associated with a specific accelerator. If multiple or different accelerators are present, they would have different values in this field. Figure~\ref{fig:new-trait}-L2 to L4 show the available parameters. The code generated for the DMA initialization is executed by the CPU only once per application.

\noindent \lstinline{- init_opcodes}: defines a flow of opcodes that should be sent to initialize or reset the accelerator for a new kernel execution. During application runtime, these opcodes are sent $N$ times, where $N$ is the number of kernels in an application that can be mapped onto the custom accelerator. In Figure~\ref{fig:new-trait}-L7, we define that the reset opcode must be included to support the described accelerator. The opcode's functionality is derived from the \lstinline{opcode_map} parameter below. 

\noindent \lstinline{- accel_dim}: defines the size of the accelerator for each dimension of the implemented algorithm. Figure~\ref{fig:new-trait}-L9 shows an example, specifying that the \textit{accelerator} supports a tiled MatMul$_{4x4x4}$ version of the implemented algorithm.

\noindent \lstinline{- permutation_map}: defines the order in which nested loops execute. In Figure~\ref{fig:new-trait}-L12, we switch the order of the two innermost loops, potentially enabling the data structure that uses \textit{[m,k]} indices to be stationary, as the other data structures are streamed in/out of the accelerator. In our MatMul example (Figure~\ref{fig:lst-mlir-matmul-scf}), this enables an \lstinline{A} stationary dataflow (Figure~\ref{fig:matmul-with-accel}).

\begin{figure}[t]
\centering
\begin{adjustbox}{minipage=\linewidth,scale=0.95}
\centering
\input{lsts/opcode_map_syntax}
\caption{Opcode Map Syntax. A dictionary for accelerator opcodes and actions.}\label{fig:opcode_map_syntax}
\end{adjustbox}
\end{figure}

\begin{figure}[t]
\centering
\begin{adjustbox}{minipage=\linewidth,scale=0.95}
\centering
\input{lsts/opcode_flow_syntax}
\caption{Opcode Flow Syntax. The sequence of opcodes to implement a specific dataflow of host-accelerator communication.}\label{fig:opcode_flow_syntax}
\end{adjustbox}
\end{figure}

\noindent \lstinline{- opcode_map}:
describes accelerator opcodes as key-value pairs. Following the syntax scheme shown in Figure~\ref{fig:opcode_map_syntax}, the key, or \textit{opcode\_entry}, is an identifier that maps to a list of actions, or \textit{opcode\_list}, which represents sequential memory operations that have to be performed to drive the accelerator. 
Each action, or \textit{opcode\_expr} (\lstinline{send}, \lstinline{send_literal}, \lstinline{send_dim}, \lstinline{send_idx}, \lstinline{recv}), implements different types of copies to/from the DMA memory-mapped region. The \lstinline{send} and \lstinline{recv} actions take an input. The input is a number that is used to represent one of the arguments to the \mbox{\lstinline{linalg.generic}} operation, e.g., 0, 1, or 2 would map to A, B, or C, respectively, in the MatMul example (Figure~\ref{fig:lst-mlir-matmul-generic}-L12-13). During code generation, this information is used to copy the needed tile to the memory-mapped region. For example, Figure~\ref{fig:new-trait}-L15 shows an opcode with identifier \textit{``sA''} that issues copies \textit{to} the accelerator for the \textit{literal} value \texttt{0x22} and then for the \textit{data} associated with the tile of argument $0$. Furthermore, \lstinline{send_dim} and \lstinline{send_idx} can be used to send tile dimensions or tile indices, which could be used to drive more complex accelerators. Subsequent text will refer to an \textit{opcode\_entry}, such as \textit{``sA''}, simply as \textit{opcode.}

\noindent \lstinline{- opcode_flow}: represents valid opcode/data transfer \textit{flows} and respects the syntax scheme shown in Figure~\ref{fig:opcode_flow_syntax}. Figure~\ref{fig:new-trait}-L23 shows an example, which defines an \textit{input A stationary} (associated with argument $0$) valid flow implemented with two opcodes, using the identifiers defined in the \mbox{\lstinline{opcode_map}.}
Additional valid examples for \textit{output C stationary} and \textit{nothing stationary} flows are shown in lines 24 and 25 of Figure~\ref{fig:new-trait}.
The information in \lstinline{opcode_flow} is parsed and the set of parentheses is understood as a proxy to specify multiple scopes for sequential or nested \textit{for} loops in the algorithm. Following this flow, logic related to \textit{``sA''} would be transmitted inside of the second loop (Figure~\ref{fig:matmul-with-accel}-L8 to L10), and logic related to \textit{``sBcCrC''} would appear in the innermost loop (Figure~\ref{fig:matmul-with-accel}-L12 to L18). 
Suppose the user decides to forego the opportunity to specify \textit{input A} as stationary, then the opcode flow could become \textit{``(sA~sB~cC~rC)''}, and all communication driver logic would be generated in the innermost loop.

\begin{figure}[t]
\centering
\includegraphics[width=1\linewidth]{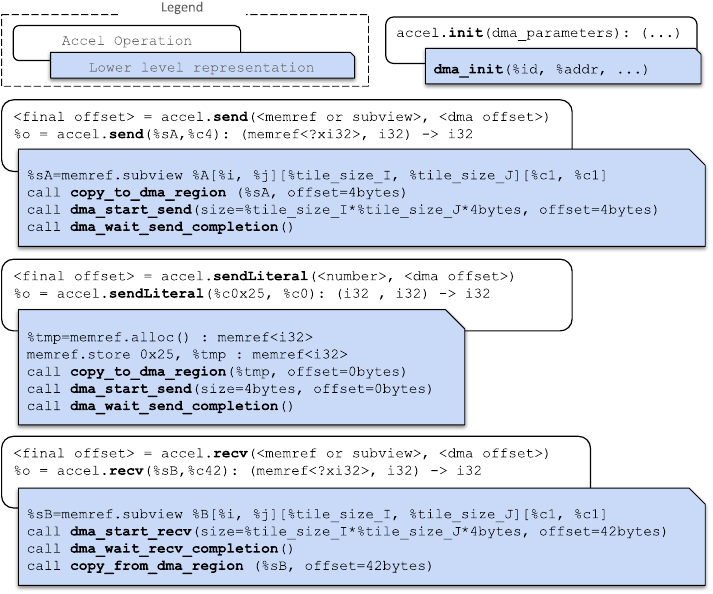}
\caption{\added{\newt{Semantics and lowering of \lstinline{accel} operations.}}}
\label{fig:accel-diagram}
\end{figure}

\noindent\textbf{The \lstinline{accel} dialect:} %
Before generating function calls for \emph{runtime replacement} to the DMA runtime library (described in Section~\ref{sec:dma-rt-lib}), we perform \emph{host code transformations}~\circled[5] (Figure~\ref{fig:compiler-flow}) by lowering the \lstinline{linalg.generic} operation, with the proposed \lstinline{trait}, to standard MLIR dialects (\lstinline{scf}, \lstinline{arith}, \lstinline{memref}) and a new dialect that we call \lstinline{accel}. Operations in the \lstinline{accel} dialect abstract host-accelerator transactions, such as initialization, memory transfers, and synchronization. 
\added{Figure~\ref{fig:accel-diagram} presents the core \lstinline{accel} operations and their semantics, providing examples of how these operations map onto our custom AXI DMA library calls. Additionally, Figure~\ref{fig:matmul-with-accel} shows how the \lstinline{accel} operations are used in our MatMul example.}

Note that it is easier to perform analysis and transformations of operations when they are expressed in our \lstinline{accel} dialect, as opposed to using a lower-level abstraction. With lower-level abstractions such as \lstinline{llvm}, function calls and additional logic have already been exposed: additional instructions must be present in the IR to implement buffer slicing, size/offset calculations, and function calls to copy data to/from the DMA regions. Performing analysis and transformations in the \lstinline{llvm} abstraction is more challenging, as traversal of control flow blocks and LLVM instructions are necessary. Instead, operations in the intermediate \lstinline{accel} dialect encode the relevant information, and are easily relocated during transformation passes, respecting dependencies without requiring complex compiler analysis. This approach facilitates implementing communication flows that consider one of the data structures to be stationary by simply hoisting the \lstinline{accel} operations up to the right loop nest level, while considering the flow patterns.
Finally, the \lstinline{accel} dialect provides an intermediate step before runtime call replacement. In this work we target our AXI DMA runtime library described in Section~\ref{sec:dma-rt-lib}, but further extensions could implement the transformation of \lstinline{accel} operations into other runtime libraries such as OpenCL~\cite{stone2010opencl} or SYCL~\cite{reyes2020sycl}, which are commonly used to interface with SoC FPGA accelerators.

\section{Experiments and Results}\label{sec:results}

To evaluate AXI4MLIR, we use a PYNQ-Z2 board that includes a Zynq-7000 SoC with a dual-core ARM Cortex-A9 CPU~(650 MHz), and a library of tile-based accelerators derived from SECDA-TFLite~\cite{haris2023secda-tflite} implemented with AXI-S interface and opcodes with a micro-ISA. For workloads, we target a suite of kernels covering a range of dimensions, as well as an end-to-end machine learning application. 
We leverage hand-written baselines, which we discuss in Section~\ref{subsec:exper:baseline}.
Section~\ref{subsec:exper:matmul} evaluates accelerators implementing MatMul, comparing inference performance against a hand-written baseline, identifying potential bottlenecks, and showcasing the benefits of our optimized dataflows. Section~\ref{subsec:exper:matmul_flex} highlights the value of AXI4MLIR by demonstrating how to handle accelerators with configurable parameters such as tile sizes and dataflows. We showcase how to use AXI4MLIR with a convolution-based accelerator in Section~\ref{subsec:exper:conv}. Finally, Section~\ref{subsec:exper:e2e} shows how AXI4MLIR can work in the context of a complete application, evaluating the TinyBERT model~\cite{jiao2019tinybert}.

\subsection{Hand-written Baselines}
\label{subsec:exper:baseline}

The next experiments employ hand-written optimized driver code derived from the SECDA-TFLite accelerator toolkit~\cite{haris2023secda-tflite} to establish performance baselines. SECDA-TFLite presents a state-of-the-art toolchain and methodology for HW/SW co-design of embedded machine learning accelerators targeting FPGA SoC devices. With host-driver code written in C{++}, these manual baselines will be labeled as \textit{cpp\_MANUAL}. All baselines are implemented with various tiling strategies, with no additional data transfer overheads and with the fewest number of data transfer calls for the selected dataflow.

\subsection{Matrix-Multiplication Experiments}
\label{subsec:exper:matmul}

The tile-based accelerators used here resemble vector MAC engines~\cite{albericio2016cnvlutin,zhang2016cambricon,chen2014dadiannao,zhang2015optimizing} implementing MatMul algorithms.
They vary in input/output buffer size and supported dataflow. From the CPU-host perspective, some of them can support varying degrees of data reuse when the appropriate opcode stream drives the accelerator. Table~\ref{tab:accelerators} presents a short summary of their functionality, where \textit{size} stands for the supported tile size of the accelerator. For example, $\mathrm{v1_{4}}$ is a  MatMul$_{4x4x4}$ accelerator that does not support data reuse and only supports $tM,tN,tK==4,4,4$ tiles. For $\mathrm{v1_{4}}$, AXI4MLIR will tile the algorithm's loops in the host code, taking into account the accelerator size of 4 and all the data movement will happen in the innermost loop - ``\mbox{\lstinline{opcode_flow <(sA sB cCrC)>}}''. For $\mathrm{v2_{8}}$, AXI4MLIR will tile the computation by 8 and generate code to maximize the reuse of one of the inputs. In $\mathrm{v2}$, a stationary (As) is implemented with \mbox{\lstinline{opcode_flow <(sA (sB cCrC))>}}.

Accelerators $\mathrm{v3}$ and $\mathrm{v4}$ can also reuse their output data structures. Accelerator $\mathrm{v4}$, marked with \textit{flex size}, supports computations of non-square tiles, i.e., $\mathrm{v4_{16}}$ can process a MatMul of $tM,tN,tK=32,16,64$, as long as $tM,tN,tK$ are divisible by $16$ and fit in the accelerator's memory. 
All accelerators were implemented using HLS pipelining and unrolling to maximize the number of internal processing elements instantiated and their arithmetic throughput.
The last column of Table~\ref{tab:accelerators} reports throughput \mbox{(OPs/cycle)} for each accelerator, highlighting that many arithmetic operations are executed in parallel at each cycle. Different types of accelerators with the same size have the same throughput, and accelerators with bigger sizes provide higher throughput.
All bar graphs presented in this section represent the average of 5 independent runs with the same configuration.

\begin{figure}[t]
\centering
\includegraphics[width=1\linewidth]{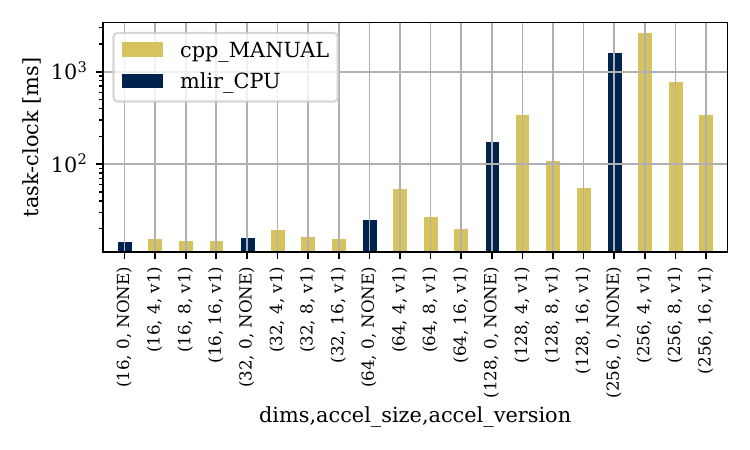}
\caption{Runtime characterization CPU vs. Accelerator execution for Matrix Multiplication problems. Note how an accelerator only becomes relevant for problems with $dims\geq64$ and $accel\_size\geq8$.}
\label{fig:exp-manual-vs-mlir}
\end{figure}

\noindent\textbf{Accelerator relevance.}
In order to evaluate the performance of the accelerators defined in Table~\ref{tab:accelerators}, we conducted experiments to compare the runtime of the CPU execution (\textit{mlir\_CPU}) against the manual C{++} implementation (referred to as \textit{cpp} for short) of the driver code using the accelerators. The task clock was used as a metric to measure the execution time of the benchmarks. We present the results of the experiments in Figure~\ref{fig:exp-manual-vs-mlir}, which plots the task clock on the y-axis (smaller is better) and only includes the ``Nothing Stationary flow'', which means that the data transfers happen in the innermost loop.

Looking at Figure~\ref{fig:exp-manual-vs-mlir}, we can see that the accelerator offload only becomes relevant (i.e., executes faster than the CPU) for problems with $dims\geq64$, where $dims=M=N=K$. For problems with smaller dimensions, CPU execution will be faster than the accelerator. In addition, the results in Figure~\ref{fig:exp-manual-vs-mlir} suggest that accelerators only become relevant if \mbox{$accel\_size=tM=tN=tK\geq8$}. For smaller accelerator sizes, the CPU execution is faster than the accelerator.

These observations suggest that the performance benefits of using the accelerators are limited for ranges of problem sizes and accelerator sizes. Therefore, it is important to carefully choose the appropriate accelerator configuration for a given problem to achieve the best performance. Consequently, for the next experiments we will limit our focus to problems with $dims\geq64$ and accelerators with $accel\_size\geq8$.

\begin{table}[t]
\caption{Accelerators used in the experiments. Synthesized with AMD/Xilinx Vitis at 200MHz.}\label{tab:accelerators}
\resizebox{1\linewidth}{!}{%
\begin{tabular}{ccccc}
\cline{1-3}\cline{5-5} 
\multicolumn{1}{|c}{\textbf{Type}} & \multicolumn{1}{c}{\textbf{Possible Reuse}} & \multicolumn{1}{c|}{\textbf{Opcode(s)}} & & \multicolumn{1}{|c|}{\textbf{Configurations}} \\ \cline{1-3}\cline{5-5} 
\textbf{$\mathrm{v1_{size}}$}               & Nothing                                     & sAsBcCrC & & \textbf{(Size, OPs/Cycle)}                   \\\cline{5-5}  
\textbf{$\mathrm{v2_{size}}$}               & Inputs                                      & sA, sB, cCrC & & \multirow{1}{*}{(4, 10)}                 \\
\textbf{$\mathrm{v3_{size}}$}               & Ins/Out                                    & sA, sB, cC, rC & & \multirow{1}{*}{(8, 60)}                \\
\textbf{$\mathrm{v4_{size}}$}          & Ins/Out (flex size)                                    & sA, sB, cC, rC & & \multirow{1}{*}{(16, 112)}             \\ \cline{1-3}\cline{5-5} 
\end{tabular}%
}%
\end{table}

\noindent\textbf{AXI4MLIR generated vs. Manual implementation.}

\begin{figure}[t]
\centering
\includegraphics[width=1\linewidth]{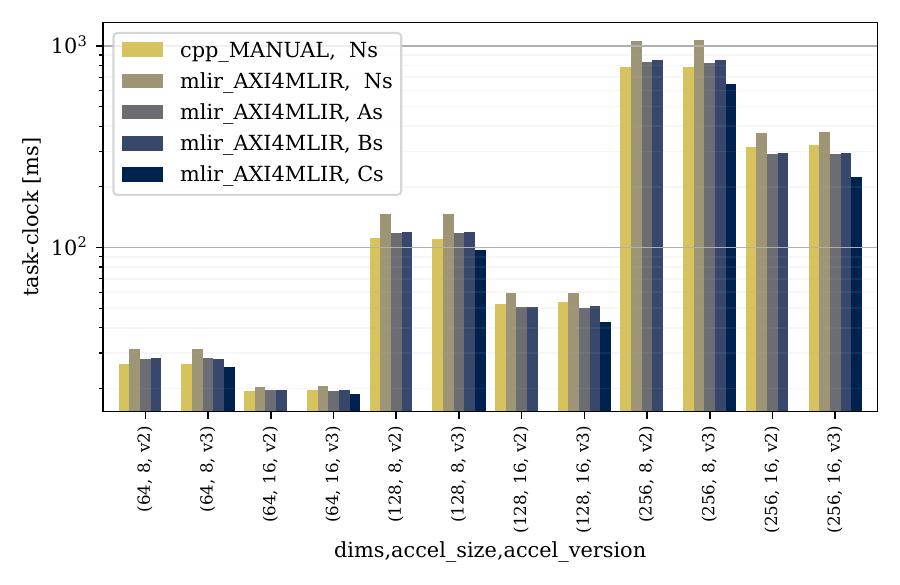}
\caption{Runtime results on Matrix Multiplication kernels. Manual implementation of Ns flow vs. AXI4MLIR generated driver code for different flow strategies, Ns, As, Bs, Cs. All bar groups follow similar trends. Ns, As, and Bs \textbf{bottlenecks are analysed and addressed} in following experiments.}
\label{fig:exp-manual-vs-mlir-genereated-all-baseline}
\end{figure}

AXI4MLIR provides several benefits. First, our passes automatically tile data mapped to the CPU memory hierarchy, leveraging spatial and temporal locality. The second benefit is the ability to automatically generate specific flows, such as the Nothing Stationary (Ns) flow, which can be tedious and error-prone when done manually. Additionally, AXI4MLIR provides an efficient path to flow strategies that can potentially improve performance, such as input A or B stationary (As, Bs) flows. Figure~\ref{fig:exp-manual-vs-mlir-genereated-all-baseline} presents these results.

First, we compare the differences in execution time between a \textit{manual implementation} (see Section~\ref{subsec:exper:baseline}) of an Ns flow strategy and an \textit{AXI4MLIR generated} Ns flow strategy, represented by the first two bars in each group of bars in Figure~\ref{fig:exp-manual-vs-mlir-genereated-all-baseline}. The remaining bars in each group of bars show results for automatically generated flow strategies, with As and Bs for $\mathrm{v2}$ accelerators and As, Bs, and Cs for $\mathrm{v3}$ accelerators. Looking at Figure~\ref{fig:exp-manual-vs-mlir-genereated-all-baseline} we see that some flows, especially Cs, provide improvements. 
To achieve this, the user simply has to encode the information for Cs (or other flows) during compilation. 
For example, we can encode Cs using the opcode\_flow previously presented in Figure~\ref{fig:new-trait}-L25 in the the operation's trait.

Next, in Figure~\ref{fig:exp-manual-vs-mlir-genereated-all-baseline}, we focus on the results with the $\mathrm{v3}$ accelerator. Here, we see that AXI4MLIR generated Cs performs better than the manually generated Ns, although the other flows are not performing as expected. First, we would expect the performance of AXI4MLIR generated Ns to have similar/closer task clock performance than manual Ns. And second, we would also expect As and Bs flows to always outperform Ns due to the degree of reuse, as they copy less data and can keep the accelerator better utilized.
Hence, this first implementation has room for improvement and, in the following experiment, we \textit{identified and fixed the bottlenecks} by analyzing performance counters and implementing optimizations that specialize memory copies.

\noindent\textbf{Identifying bottlenecks \& improving AXI4MLIR codegen.}
Next, we identify performance bottlenecks in AXI4MLIR generated copies and improve upon them to enhance the performance of the workloads when using the custom hardware accelerators. Specifically, the experiment compares the performance of manually implemented host-accelerator driver code with AXI4MLIR generated code for Ns, As, Bs, and Cs flows in terms of branch-instructions, cache reference counters, and the task clock. These metrics were obtained using the \lstinline{perf} tool~\cite{perfteamNDperf} to profile the application and retrieve counters for CPU perf\_events over 5 runs.

\begin{figure}[t]
\begin{subfigure}[t]{1\columnwidth}
\centering
\includegraphics[width=1\linewidth]{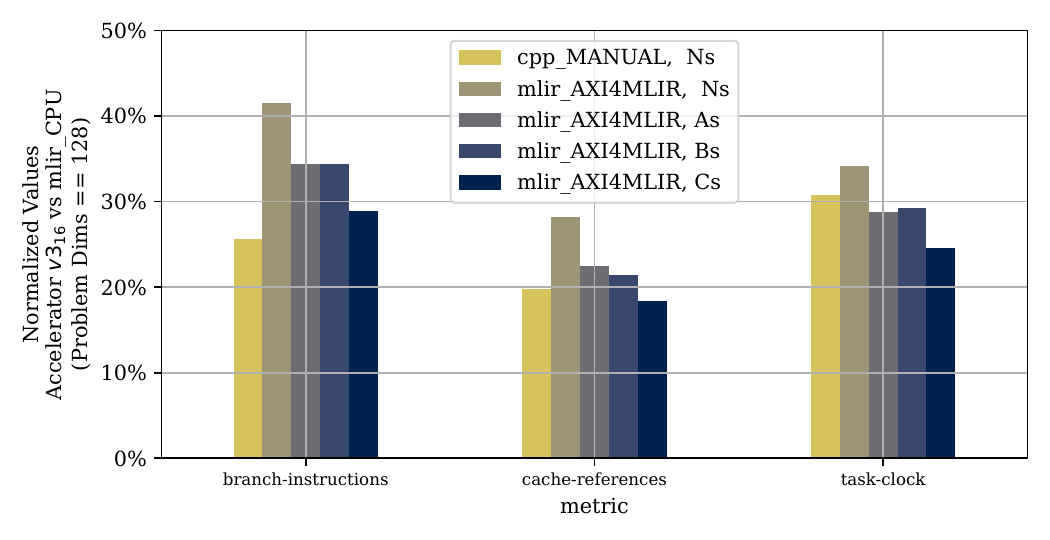}
\caption{Without the MemRef-DMA buffer copy optimization. Generated host-accelerator code has overheads if not specialized.}\label{fig:exp-manual-vs-mlir-counters-baseleline}
\end{subfigure}

\begin{subfigure}[t]{1\columnwidth}
\centering
\includegraphics[width=1\linewidth]{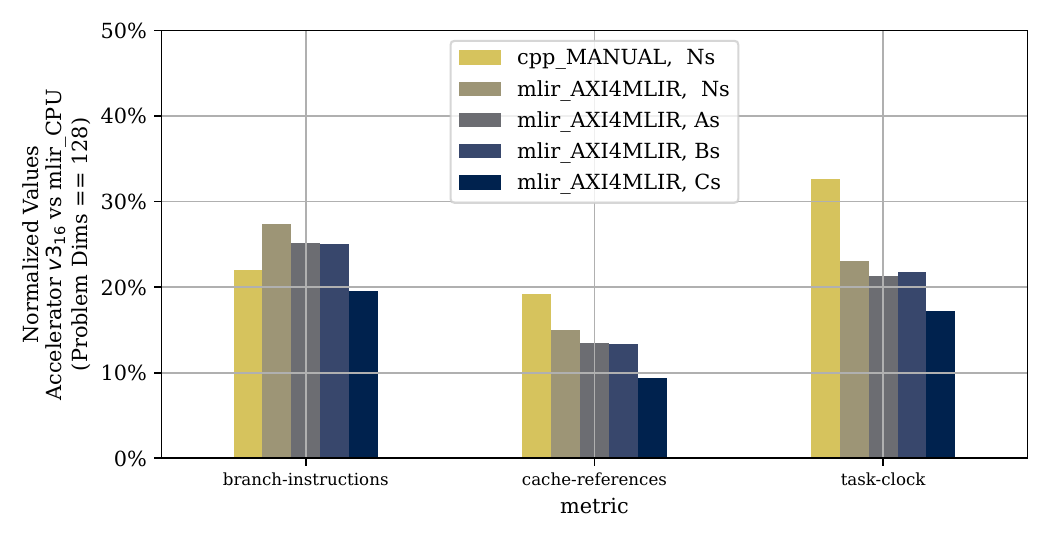}
\caption{With MemRef-DMA buffer copy optimization. AXI4MLIR improves accelerator performance over manual Ns implementation.}\label{fig:exp-manual-vs-mlir-counters-optimized}
\end{subfigure}
\caption{Cache, branch, and runtime metrics of different tools and strategies using $\mathrm{v3_{16}}$ accelerator with problem size ($dims==128$). \added{Normalized values against CPU (without accelerator) executions of same problem size.}
}\label{fig:exp-manual-vs-mlir-counters-comparison}
\end{figure}

Figure~\ref{fig:exp-manual-vs-mlir-counters-baseleline} shows branch instructions, cache reference counters, and the task clock for $dims==128$, for the $\mathrm{v3_{16}}$ accelerator that supports input and output stationary flows. The trends are similar to other problem and accelerator sizes. Our results are normalized to the same counters collected on a CPU-only execution of the same problem size. In each group we show results for AXI4MLIR automatically generated code for Ns, As, Bs, Cs flows, and compare against manual implementations (first bar of a group) for copying the necessary data through the DMA memory-mapped region. MLIR applications have to consider MLIR memory references (presented in Section~\ref{sec:mlir-memrefs}), but manual implementations use bare C-arrays. To support generality, MLIR copies between MemRef and the raw array (DMA buffer region) are implemented with a recursive call, loading and storing one element at a time. This is necessary to support $rank=N$ MemRefs, where strides in all dimensions are different from 1.

\begin{figure*}[t]
\centering
\includegraphics[width=1\linewidth]{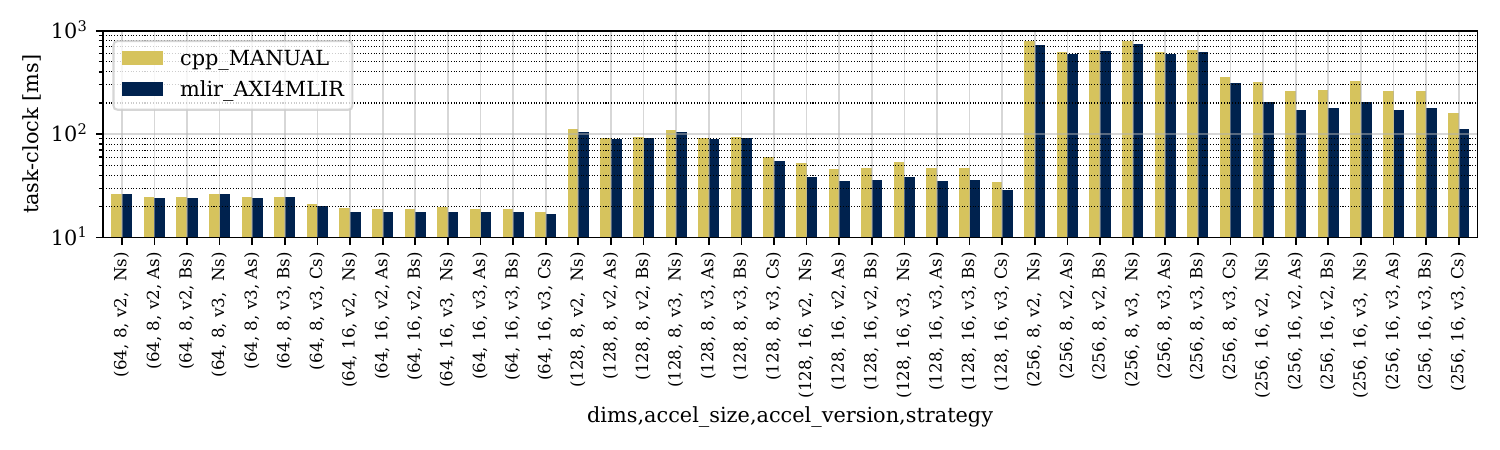}
\caption{Runtime comparison of manual implementation of driver code and AXIMLIR generated. Each set of two bars have a matching Accelerator Type, Accelerator Size, and Flow Strategy (Ns, As, Bs, Cs). AXI4MLIR is better in all cases.}\label{fig:exp-manual-vs-mlir-genereated-all-best_mr_copy}
\end{figure*}

In order to address this issue, we implemented an optimization for when $strides[N-1]==1$ (i.e., elements in $N-1$ dimension are adjacent to each other in memory) and specialized MemRef copies for some known rank sizes, such as $rank==2$. For this scenario, we leverage the spatial locality and implement the copy not with individual load and store instructions, but by calling \mbox{\lstinline{std::memcpy(src, dst, size)}}. When compiling this function for our platform, the compiler will inline the assembly, implementing a vectorized copy that improves the performance of the copy operation. The implications of this optimization are twofold. First, it reduces the number of branch references because there is no need for branching to handle non-unitary strides or to iterate over an arbitrary number of dimensions, resulting in better control flow and branch prediction. Second, the vectorized code reduces the number of cache references because the data is accessed sequentially in memory. 
\added{Therefore, there will only be two cache reference to fetch the cache line containing the requested data, and subsequent loads within the same cache line will not require additional cache references as they are read from the vector VFP registers~\cite{arm2023neonregisters}.}
The results for this optimization are presented in Figure~\ref{fig:exp-manual-vs-mlir-counters-optimized}.

After incorporating this optimization, the AXI4MLIR generated driver code executed faster on all accelerators as compared to their respective manual implementations. In Figure~\ref{fig:exp-manual-vs-mlir-genereated-all-best_mr_copy}, we compare AXI4MLIR against manual implementations for Ns, As, Bs, and Cs, and found that the compiled generated driver code provided by AXI4MLIR is consistently faster (1.18$\times$ average speedup and 1.65$\times$ max speedup), thanks to its ability to leverage proper tiling for the CPU's memory hierarchy, resulting in a 10\% average and 56\% max reduction in cache references.

\subsection{Matrix-Multiplication with flexible sizes}
\label{subsec:exper:matmul_flex}

\begin{figure}[t]
\centering
\includegraphics[width=1\linewidth]{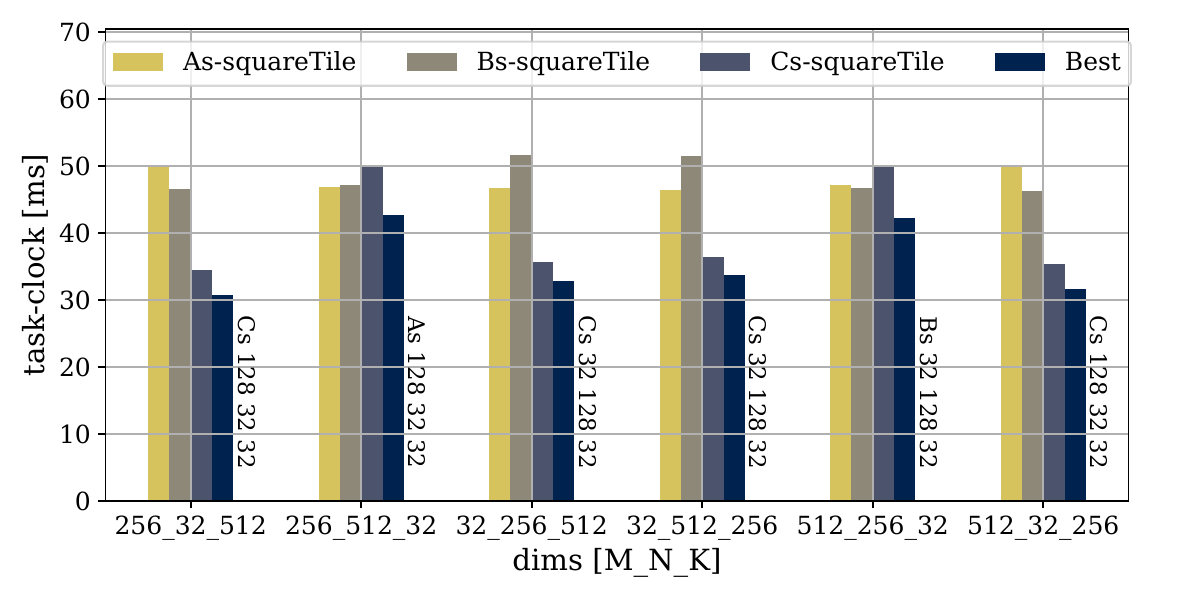}
\caption{MatMul problem permutations (v4 accelerator) for different strategies. For the ``Best'' strategies we annotate the chosen flow and tiling values.}\label{fig:v4_graph}
\end{figure}

Runtime configurable accelerators allow for fine-grained hardware tuning for specific problems. With AXI4MLIR, we can generate host code to configure and optimize flexible accelerators for the target problem. To demonstrate this capability, we evaluate multiple permutations of a MatMul problem on the $\mathrm{v4}$ accelerator. The $\mathrm{v4}$ accelerator supports multiple dataflow strategies and adjustable tile sizes for its $tM$, $tN$, and $tK$ dimensions. The intuition is that scientific and machine learning workloads present problem sizes with different values for each dimension, sometimes resulting in tall/skinny matrices during execution. Tiling the problem in the accelerator with different dimensions for $tM$, $tN$, and $tK$, and selecting the appropriate flow strategy can be beneficial for the application. 

When using AXI4MLIR, a developer is \textit{not} limited to one configuration of an accelerator. Based on the user's knowledge of the application, AXI4MLIR can automatically generate the driver for accelerators with adjustable dimensions. This flexibility allows for a more thorough exploration of the design space, enabling the developer to find the best sizes for $tM$, $tN$, $tK$, and the best flow strategy for each problem instance.

In Figure~\ref{fig:v4_graph}, we compare four different heuristics and use them to choose the best tiling and dataflow configuration for a MatMul problem. We evaluate performance in terms of execution time. We profile the problem with $M$,$N$, and $K$ dimensions permuted from the following values: $[32,256,512]$. Hence, the theoretical minimum number of multiply-accumulate operations required for all permutations is the same. Here, the \emph{As-squareTile}, \emph{Bs-squareTile}, and \emph{Cs-squareTile} heuristics try to find the best configuration to reduce the total memory access count given the constraint of tiling the MatMul with square tiles (i.e., $tM=tN=tK=T$), with A, B, and C stationary dataflow, respectively. The fourth heuristic, \emph{Best}, chooses between all dataflows and flexible tiling options, only sharing the choice of the accelerator. In Figure~\ref{fig:v4_graph} we annotate the ``\emph{Best}'' configuration found for each problem. 

\noindent\textbf{Square tiling.}
We observe that as we change the problem permutation, the best flow between \emph{As-squareTile}, \emph{Bs-squareTile}, and \emph{Cs-squareTile} tiling strategies changes. The best flow depends on the problem shape, the size, and the available accelerator buffer space. $T=32$ was selected for all square flows because it is the biggest value, so the tiles fit inside the accelerator's internal memory.

\noindent\textbf{Flexible tiling.}
The \emph{Best} heuristic, selected from non-square strategies, outperforms square tiling by leveraging flexible tiling sizes. AXI4MLIR can generate code to utilize larger tile sizes in various dimensions, taking advantage of the $\mathrm{v4}$ accelerator's unrestricted tiling factors and improving the accelerator's internal memory utilization.

\noindent\textbf{Configurations.}
Manually implementing all configurations' driver code for even a simple accelerator such as $\mathrm{v4}$ is very time-consuming. AXI4MLIR can quickly generate hostcode for configurable accelerators easily, enabling the developer to specify an accelerator configuration per problem instance.

\begin{figure}[htbp]
\centering
\begin{adjustbox}{minipage=\linewidth,scale=0.95}
\begin{subfigure}[t]{1\textwidth}
\centering
\begin{lstlisting}[style=customMLIR,language=mlir]
accel_dim = map<(B,H,W, iC,oC,fH,fW) -> 
                (0,0,0,256, 1, 3, 3)>, // Tiling
opcode_map< 
  sIcO=[send_literal(70), send(0)], // send 3D input window 
                                    //  and compute
  sF=[send_literal(1), send(1)],    // send 3D filter
  rO=[send_literal(8), recv(2)],    // recv 2D output slice
  rst=[send_literal(32), send_dim(1,3),  // set filter size 
       send_literal(16), send_dim(0,1)]> // set iC size
opcode_flow <(sF (sIcO) rO)> // filter+output stationary
init_opcodes <(rst)>  
\end{lstlisting}
\caption{Opcode Map and Flow for Conv2D accelerator.}\label{fig:conv_attributes}
\end{subfigure}

\begin{subfigure}[t]{1\textwidth}
\centering
\input{lsts/conv-with-accelops}
\caption{\newt{IR to drive the Conv2D accelerator with an output-stationary flow.}}\label{fig:conv-with-accel}
\end{subfigure}
\end{adjustbox}
\caption{\added{Information added to the linalg.generic traits to capture convolution accelerator behavior in MLIR and IR with accel operations.}}
\label{fig:mlir-conv-kernel-traits}
\end{figure}

\subsection{Convolution}
\label{subsec:exper:conv}

We show the flexibility of AXI4MLIR by generating driver code for a convolution-based accelerator executing different problems sizes. This accelerator supports varying input channel (\textit{iC}) and filter (\textit{fHW}) sizes, computing one output slice (all elements in one output channel - \textit{oC}) per iteration. To orchestrate the execution, multiple instructions have to be sent to the accelerator. 
\added{
This orchestration is achieved by compiling the driver code derived from the MLIR \lstinline{accel} code (Figure~\ref{fig:conv-with-accel}). The \lstinline{accel} code is generated after a transformation pass takes into account the attributes shown in Figure~\ref{fig:conv_attributes} and MLIR's \lstinline{linalg.conv_2d_nchw_fchw} operations. Note that if the convolution operation has \textit{iC}, \textit{fH}, \textit{fW} dimensions that are smaller than the dimensions in \lstinline{accel\_dim}, no tiling will be performed across these dimensions. In the convolution example (Figure~\ref{fig:mlir-conv-kernel-traits}), upon accelerator reset, we use \lstinline{send\_dim(1,3)} to send to the accelerator the filter size as the dimension `3' of data structure `1' (i.e., the filter), and we use \lstinline{send\_dim(0,1)} to send the input channel size as the dimension `1' of the data structure `0' (i.e., the input).
}

\begin{figure}[t]
\centering
\includegraphics[width=1\linewidth]{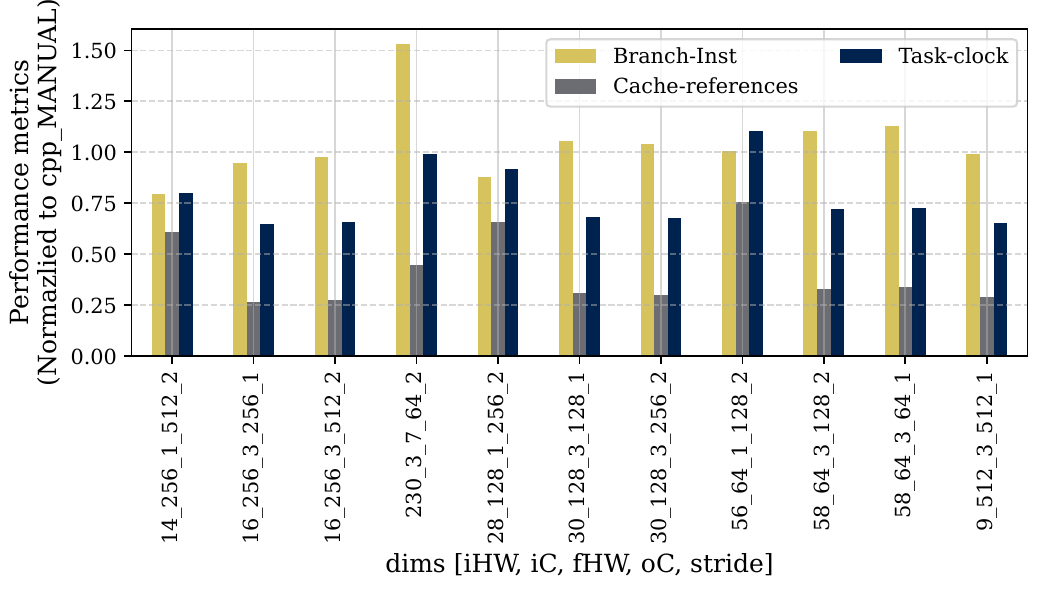}
\caption{ResNet18 convolution layers: AXI4MLIR vs. Manual.}\label{fig:conv_graph}
\end{figure}

We evaluate the performance of AXI4MLIR during the execution of all convolution layers of ResNet18~\cite{wang2017resnet}.
Figure~\ref{fig:conv_graph} presents performance metrics normalized to the runtime of layer-specific manual C++ driver code. The results observed here present similar trends as observed in the MatMul experiments. Only one layer (\textit{56\_64\_1\_128\_2}) presented a 10\% slowdown, contrary to previous trends. The slowdown happened because \textit{fHW (1)} and \textit{iC (64)} were too small, and the overhead of dealing with small MemRefs was not overcome since we could not leverage the \textit{strided copy optimization} presented in Section~\ref{subsec:exper:matmul}. Smaller AXI4MLIR speedups are observed every time that \textit{fHW}$\mathbf{==1}$. That said, AXI4MLIR achieves better runtime performance on 10 out of 11 ResNet18 layers, with 1.28$\times$ and 1.54$\times$ average and max speedup, respectively, thanks to the improved CPU cache performance.

\subsection{End-To-End Analysis}
\label{subsec:exper:e2e}

Finally, we evaluate AXI4MLIR when compiling a natural language processing model to co-execute on both the CPU and the $\mathrm{v4_{16}}$ accelerator. We benchmark TinyBERT~\cite{jiao2019tinybert}, a compact transformer~\cite{wolf2020transformers} model for Masked Language Modeling and Next Sentence Prediction targeted at mobile and embedded devices. We translate TinyBERT to MLIR IR using Torch-MLIR~\cite{torchmlir2021torchmlir} and compare the inference performance of CPU execution (using -O3 during compilation) against co-execution using the \textit{``Ns''} offloading approach and the \textit{``Best''} approach, which employs the heuristics presented in Section~\ref{subsec:exper:matmul_flex}. 

\begin{figure}[htbp]
\centering
\includegraphics[width=1\linewidth]{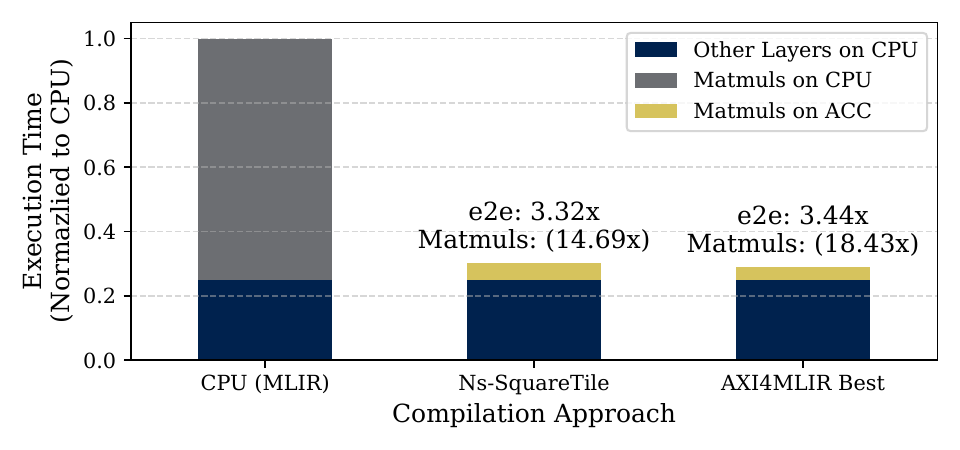}
\caption{Execution time of the TinyBERT model with $\mathrm{batch\_size}==2$. Each bar represents a compilation strategy. Speedups for end-to-end ($e2e$) and for accelerated MatMul layers are shown as annotations.}\label{fig:tinybert}
\end{figure}

As we can see in Figure~\ref{fig:tinybert}, AXI4MLIR achieves a 3.4$\times$ speedup in end-to-end execution, with an 18.4$\times$ speedup in the accelerated MatMul layers that represent 75\% of the original CPU runtime. This experiment showcases how AXI4MLIR can be used during evaluation and optimization of natural language processing models on embedded devices. Our study highlights that developers can easily co-design the accelerators when targeting full workloads, which enables efficient exploration and utilization of both CPU and accelerator resources.

\section{Related Work}\label{sec:rw}

Prior studies~\cite{Sozzo2018Frost,loncar2020hls4ml,Zhang2020dnnexplorer,tvm2020vta,ye2020HybridDNN,agostini2022SODA,agostini2022sodaopt,stjerngren2022Bifrost} have proposed new accelerator designs or presented new methodologies to generate flexible accelerators for a wide range of algorithms. However, these approaches fall short in providing insights into how host-to-accelerator transfers should be performed. Most of these tools assume that the required data is already placed in the accelerator's internal buffers. There have also been efforts to support hardware/software co-design of an accelerator for an application~\cite{Skalicky2018Hot,agostini2020tflitesoc,haris2023secda-tflite}. However, these implementations adopt a simple offload model, where execution of the kernel code is simply \textit{replaced} by runtime calls that copy the data to-and-from the accelerator, \textit{without considering the host memory hierarchy or accelerator features}, which would require manual driver code modifications.

HeteroFlow~\cite{xiang2022heteroflow}, an FPGA accelerator programming model, decouples algorithm specification from data placement optimization using a new primitive ``\lstinline{.to()}''. This approach exposes data placement specification at various granularities, achieving efficient code generation while matching optimized manual HLS designs. HeteroFlow does not support arbitrary custom accelerators, as it is limited to accelerators co-designed with its framework (extended HeteroCL~\cite{lai2019heterocl}). It also requires the new primitive to be used while describing the algorithm in Python, imposing manual application modification. Unlike HeteroFlow, AXI4MLIR utilizes MLIR to target languages employing \lstinline{linalg.generic} operations during compilation, eliminating the need for manual transformation.

Several other studies have addressed the challenge of efficiently mapping algorithms and their loops onto accelerators through operation scheduling. Notably, Interstellar~\cite{yang2020interstellar}, DMazeRunner~\cite{dave2019dmazerunner}, and PolySA~\cite{cong2018polysa} delve into more versatile loop structures by adopting diverse loop representations for DNN layers. CoSA~\cite{huang2021cosa} and Vaidya et al.~\cite{vaidya2022sched} tackle the generation of execution schedules for DNNs in a time-efficient manner, leveraging constrained optimization solvers. Self-tuning algorithms have also been employed in addressing the scheduling problem. Approaches like ConfuciuX~\cite{kao2020confuciux}, Flex-Tensor~\cite{zheng2020flextensor}, AutoTVM~\cite{chen2018autotvm}, and Ansor~\cite{zheng2020ansor} utilize machine learning algorithms. Furthermore, Flexer~\cite{min2023flexer} employs an out-of-order scheduling technique, unbound by loop order, which orchestrates operations based on a comprehensive analysis of the data-flow graph of a given layer. Some of these works are tailored to a specific type of accelerator or algorithm.
In addition, these works primarily focus on scheduling aspects, which AXI4MLIR currently lacks as a component. Nonetheless, these scheduling techniques could potentially complement AXI4MLIR's attributes and transformations to enhance the overall accelerator integration process.

The Pattern Description Language (PDL) and Transform MLIR dialects~\cite{mlir2022transform} offer productive ways for expressing IR transformations and could be leveraged to implement similar functionality as provided by AXI4MLIR. However, PDL cannot currently identify patterns in nested MLIR regions. Additionally, the transform dialect focuses on scheduling linear algebra transformations but requires extensions for runtime call generation targeting accelerators and dataflows. In contrast, AXI4MLIR's \lstinline{opcode_map} and \lstinline{opcode_flow} extensions enable flexible automatic driver code generation for custom accelerators. Future work may involve integrating AXI4MLIR passes as Transform operations and using PDL to identify operation sequences for transformation, potentially supporting fusing operations for custom accelerator execution.

\added{
Host code generation transforms accel operations into DMA library calls. To facilitate further optimizations leveraging the MLIR infrastructure, users can modify these transformation passes while applying optimizations such as double buffering, building on our infrastructure supporting non-blocking transfers and transfer completion checks. Our ongoing work will introduce a new attribute to select inputs/outputs for double buffering. This aligns with MLIR's capability to modify and add passes to the transformation pipeline. For further efficiency, coalescing transfer requests are essential; future work will implement a transformation that consolidates multiple \lstinline{start\_send} calls into a single call after data preparation, thus reducing the need for multiple \lstinline{wait\_send} calls, which incur higher CPU accelerator-DMA synchronization costs.
}

\section{Conclusion}\label{sec:conclusion}

In this paper we presented AXI4MLIR, an extension to the MLIR compiler framework to describe AXI-based accelerators with a range of features including accelerator opcodes. We implemented attribute extensions and compiler transformations to describe and automatically generate host code that can leverage different flows of flexible accelerators, allowing us to break away from simple offload HW/SW co-design models.
After implementing data staging and accessing optimizations during communication, our results show that AXI4MLIR is effective in generating host code that efficiently uses CPU resources and accelerator features. This allows for measurable runtime improvements versus manual implementations for all tested accelerators, while providing automation and convenience during the co-design cycle.
Finally, our user-driven host code generation is entirely automated, providing a significant advantage in terms of productivity and maintainability, specially during the early stages of the co-design process.

\section*{Acknowledgment}

We acknowledge support from: the DMC Initiative, the AT SCALE Initiative, and the Compiler Frameworks and Hardware Generators to Support Innovative US Government Designs project at Pacific Northwest National Laboratory;
the Engineering and Physical Sciences Research Council (grant EP/R513222/1); 
the grant RYC2021-031966-I funded by MCIN/AEI/10.13039/501100011033 and the ``European Union NextGenerationEU/PRTR.''

\balance

\bibliographystyle{IEEEtran} %
\bibliography{bibs/mybib.bib}

\begin{thebibliography}{10}
\providecommand{\url}[1]{#1}
\csname url@samestyle\endcsname
\providecommand{\newblock}{\relax}
\providecommand{\bibinfo}[2]{#2}
\providecommand{\BIBentrySTDinterwordspacing}{\spaceskip=0pt\relax}
\providecommand{\BIBentryALTinterwordstretchfactor}{4}
\providecommand{\BIBentryALTinterwordspacing}{\spaceskip=\fontdimen2\font plus
\BIBentryALTinterwordstretchfactor\fontdimen3\font minus \fontdimen4\font\relax}
\providecommand{\BIBforeignlanguage}[2]{{%
\expandafter\ifx\csname l@#1\endcsname\relax
\typeout{** WARNING: IEEEtran.bst: No hyphenation pattern has been}%
\typeout{** loaded for the language `#1'. Using the pattern for}%
\typeout{** the default language instead.}%
\else
\language=\csname l@#1\endcsname
\fi
#2}}
\providecommand{\BIBdecl}{\relax}
\BIBdecl

\bibitem{henessy2018goldenage}
J.~Hennessy and D.~Patterson, ``A new golden age for computer architecture: Domain-specific hardware/software co-design, enhanced security, open instruction sets, and agile chip development,'' in \emph{2018 ACM/IEEE 45th Annual International Symposium on Computer Architecture (ISCA)}.\hskip 1em plus 0.5em minus 0.4em\relax Los Angeles, CA, USA: IEEE, 2018, pp. 27--29.

\bibitem{shabani2023hpca}
H.~Shabani, A.~Singh, B.~Youhana, and X.~Guo, ``Hirac: A hierarchical accelerator with sorting-based packing for spgemms in dnn applications,'' in \emph{2023 IEEE International Symposium on High-Performance Computer Architecture (HPCA)}.\hskip 1em plus 0.5em minus 0.4em\relax Montreal, QC, Canada: IEEE, 2023, pp. 247--258.

\bibitem{kim2023hpca}
B.~Kim, S.~Li, and H.~Li, ``Inca: Input-stationary dataflow at outside-the-box thinking about deep learning accelerators,'' in \emph{2023 IEEE International Symposium on High-Performance Computer Architecture (HPCA)}.\hskip 1em plus 0.5em minus 0.4em\relax Montreal, QC, Canada: IEEE, 2023, pp. 29--41.

\bibitem{zhao2022isca}
\BIBentryALTinterwordspacing
J.~Zhao, Y.~Yang, Y.~Zhang, X.~Liao, L.~Gu, L.~He, B.~He, H.~Jin, H.~Liu, X.~Jiang, and H.~Yu, ``Tdgraph: A topology-driven accelerator for high-performance streaming graph processing,'' in \emph{Proceedings of the 49th Annual International Symposium on Computer Architecture}, ser. ISCA '22.\hskip 1em plus 0.5em minus 0.4em\relax New York, NY, USA: Association for Computing Machinery, 2022, p. 116–129. [Online]. Available: \url{https://doi.org/10.1145/3470496.3527409}
\BIBentrySTDinterwordspacing

\bibitem{hsia2023asplos}
\BIBentryALTinterwordspacing
S.~Hsia, U.~Gupta, B.~Acun, N.~Ardalani, P.~Zhong, G.-Y. Wei, D.~Brooks, and C.-J. Wu, ``Mp-rec: Hardware-software co-design to enable multi-path recommendation,'' in \emph{Proceedings of the 28th ACM International Conference on Architectural Support for Programming Languages and Operating Systems, Volume 3}, ser. ASPLOS 2023.\hskip 1em plus 0.5em minus 0.4em\relax New York, NY, USA: Association for Computing Machinery, 2023, p. 449–465. [Online]. Available: \url{https://doi.org/10.1145/3582016.3582068}
\BIBentrySTDinterwordspacing

\bibitem{Zheng2022AMOSEA}
\BIBentryALTinterwordspacing
S.~Zheng, R.~Chen, A.~Wei, Y.~Jin, Q.~Han, L.~Lu, B.~Wu, X.~Li, S.~Yan, and Y.~Liang, ``Amos: Enabling automatic mapping for tensor computations on spatial accelerators with hardware abstraction,'' in \emph{Proceedings of the 49th Annual International Symposium on Computer Architecture}, ser. ISCA '22.\hskip 1em plus 0.5em minus 0.4em\relax New York, NY, USA: Association for Computing Machinery, 2022, p. 874–887. [Online]. Available: \url{https://doi.org/10.1145/3470496.3527440}
\BIBentrySTDinterwordspacing

\bibitem{munoz2023asplos}
\BIBentryALTinterwordspacing
F.~Mu\~{n}oz Mart\'{\i}nez, R.~Garg, M.~Pellauer, J.~L. Abell\'{a}n, M.~E. Acacio, and T.~Krishna, ``Flexagon: A multi-dataflow sparse-sparse matrix multiplication accelerator for efficient dnn processing,'' in \emph{Proceedings of the 28th ACM International Conference on Architectural Support for Programming Languages and Operating Systems, Volume 3}, ser. ASPLOS 2023.\hskip 1em plus 0.5em minus 0.4em\relax New York, NY, USA: Association for Computing Machinery, 2023, p. 252–265. [Online]. Available: \url{https://doi.org/10.1145/3582016.3582069}
\BIBentrySTDinterwordspacing

\bibitem{abolhasani2023sdl}
M.~Abolhasani and E.~Kumacheva, ``The rise of self-driving labs in chemical and materials sciences,'' \emph{Nature Synthesis}, vol.~0, no.~0, pp. 1--10, 2023.

\bibitem{Rao2018sdc}
\BIBentryALTinterwordspacing
Q.~Rao and J.~Frtunikj, ``Deep learning for self-driving cars: Chances and challenges,'' in \emph{2018 IEEE/ACM 1st International Workshop on Software Engineering for AI in Autonomous Systems (SEFAIAS)}, ser. SEFAIS '18.\hskip 1em plus 0.5em minus 0.4em\relax New York, NY, USA: Association for Computing Machinery, 2018, p. 35–38. [Online]. Available: \url{https://doi.org/10.1145/3194085.3194087}
\BIBentrySTDinterwordspacing

\bibitem{jumper2021alphafold}
J.~Jumper, R.~Evans, A.~Pritzel, T.~Green, M.~Figurnov, O.~Ronneberger, K.~Tunyasuvunakool, R.~Bates, A.~{\v{Z}}{\'\i}dek, A.~Potapenko \emph{et~al.}, ``Highly accurate protein structure prediction with alphafold,'' \emph{Nature}, vol. 596, no. 7873, pp. 583--589, 2021.

\bibitem{Zhang2020dnnexplorer}
X.~{Zhang}, H.~{Ye}, J.~{Wang}, Y.~{Lin}, J.~{Xiong}, W.~{Hwu}, and D.~{Chen}, ``{DNNExplorer: A Framework for Modeling and Exploring a Novel Paradigm of FPGA-based DNN Accelerator},'' in \emph{ICCAD}.\hskip 1em plus 0.5em minus 0.4em\relax New York, NY, USA: Association for Computing Machinery, 2020, pp. 1--9.

\bibitem{pengfei2020autoDNNchip}
\BIBentryALTinterwordspacing
P.~Xu, X.~Zhang, C.~Hao, Y.~Zhao, Y.~Zhang, Y.~Wang, C.~Li, Z.~Guan, D.~Chen, and Y.~Lin, ``Autodnnchip: An automated dnn chip predictor and builder for both fpgas and asics,'' in \emph{Proceedings of the 2020 ACM/SIGDA International Symposium on Field-Programmable Gate Arrays}.\hskip 1em plus 0.5em minus 0.4em\relax New York, NY, USA: Association for Computing Machinery, 2020, p. 40–50. [Online]. Available: \url{https://doi.org/10.1145/3373087.3375306}
\BIBentrySTDinterwordspacing

\bibitem{ye2020HybridDNN}
H.~Ye, X.~Zhang, Z.~Huang, G.~Chen, and D.~Chen, ``Hybriddnn: {A} framework for high-performance hybrid {DNN} accelerator design and implementation,'' in \emph{DAC}.\hskip 1em plus 0.5em minus 0.4em\relax San Francisco, CA, USA: IEEE, 2020, pp. 1--6.

\bibitem{kwon2020maestro}
H.~Kwon, P.~Chatarasi, V.~Sarkar, T.~Krishna, M.~Pellauer, and A.~Parashar, ``Maestro: A data-centric approach to understand reuse, performance, and hardware cost of dnn mappings,'' \emph{IEEE Micro}, vol.~40, no.~3, pp. 20--29, 2020.

\bibitem{gibsonDLAS2023}
P.~Gibson, J.~Cano, E.~J. Crowley, A.~Storkey, and M.~O'Boyle, ``{{DLAS: An Exploration and Assessment of the Deep Learning Acceleration Stack}},'' \emph{arXiv:2311.08909}, Nov. 2023.

\bibitem{chen2019eyeriss}
Y.-H. Chen, T.-J. Yang, J.~Emer, and V.~Sze, ``Eyeriss v2: A flexible accelerator for emerging deep neural networks on mobile devices,'' \emph{IEEE Journal on Emerging and Selected Topics in Circuits and Systems}, vol.~9, no.~2, pp. 292--308, 2019.

\bibitem{tvm2020vta}
\BIBentryALTinterwordspacing
{TVM Developers}, ``{VTA: Deep learning accelerator stack},'' 2020. [Online]. Available: \url{docs.tvm.ai/vta}
\BIBentrySTDinterwordspacing

\bibitem{loncar2020hls4ml}
J.~Ngadiuba, V.~Loncar, M.~Pierini, S.~Summers, G.~Di~Guglielmo, J.~Duarte, P.~Harris, D.~Rankin, S.~Jindariani, M.~Liu \emph{et~al.}, ``{Compressing deep neural networks on FPGAs to binary and ternary precision with hls4ml},'' \emph{ML: Science and Technology}, vol.~2, no.~1, p. 015001, 2020.

\bibitem{Skalicky2018Hot}
S.~{Skalicky}, J.~{Monson}, A.~{Schmidt}, and M.~{French}, ``{Hot \& Spicy: Improving Productivity with Python and HLS for FPGAs},'' in \emph{FCCM}.\hskip 1em plus 0.5em minus 0.4em\relax Boulder, CO, USA: IEEE, 2018, pp. 85--92.

\bibitem{agostini2020tflitesoc}
N.~Bohm~Agostini, S.~Dong, E.~Karimi, M.~T. Lapuerta, J.~Cano, J.~L. Abell{\'a}n, and D.~Kaeli, ``Design space exploration of accelerators and end-to-end dnn evaluation with tflite-soc,'' in \emph{SBAC-PAD}.\hskip 1em plus 0.5em minus 0.4em\relax Porto, Portugal: IEEE, 2020, pp. 10--19.

\bibitem{arm2003axi}
\BIBentryALTinterwordspacing
{ARM Developers}, ``{AMBA AXI and ACE Protocol Specification},'' 2020. [Online]. Available: \url{https://developer.arm.com/documentation/ihi0022/e/AMBA-AXI3-and-AXI4-Protocol-Specification}
\BIBentrySTDinterwordspacing

\bibitem{liu2018trets}
\BIBentryALTinterwordspacing
S.~Liu, H.~Fan, X.~Niu, H.-c. Ng, Y.~Chu, and W.~LUK, ``Optimizing cnn-based segmentation with deeply customized convolutional and deconvolutional architectures on fpga,'' \emph{ACM Trans. Reconfigurable Technol. Syst.}, vol.~11, no.~3, dec 2018. [Online]. Available: \url{https://doi.org/10.1145/3242900}
\BIBentrySTDinterwordspacing

\bibitem{Lattner2021mlir}
C.~{Lattner}, M.~{Amini}, U.~{Bondhugula}, A.~{Cohen}, A.~{Davis}, J.~{Pienaar}, R.~{Riddle}, T.~{Shpeisman}, N.~{Vasilache}, and O.~{Zinenko}, ``{MLIR: Scaling Compiler Infrastructure for Domain Specific Computation},'' in \emph{CGO}.\hskip 1em plus 0.5em minus 0.4em\relax Seoul, Korea (South): IEEE, 2021, pp. 2--14.

\bibitem{mlir2020linalg}
\BIBentryALTinterwordspacing
M.~Developers, ``{'linalg' Dialect},'' 2020, online accessed on 11-04-2023. [Online]. Available: \url{https://mlir.llvm.org/docs/Dialects/Linalg/}
\BIBentrySTDinterwordspacing

\bibitem{Le2020onnxmlir}
T.~D. Le, G.-T. Bercea, T.~Chen, A.~E. Eichenberger, H.~Imai, T.~Jin, K.~Kawachiya, Y.~Negishi, and K.~O'Brien, ``{Compiling ONNX Neural Network Models Using MLIR},'' \emph{ArXiv}, vol.~0, no.~0, pp. 1--8, 2020.

\bibitem{axiguide}
\BIBentryALTinterwordspacing
{Xilinx}, ``{AXI Reference Guide},'' 2012. [Online]. Available: \url{https://docs.xilinx.com/v/u/14.1-English/ug761_axi_reference_guide}
\BIBentrySTDinterwordspacing

\bibitem{rohwedder2023gpat}
C.~Salvador~Rohwedder, N.~Henderson, J.~a. P.~L. De~Carvalho, Y.~Chen, and J.~N. Amaral, ``To pack or not to pack: A generalized packing analysis and transformation,'' in \emph{Proceedings of the 21st ACM/IEEE International Symposium on Code Generation and Optimization}, ser. CGO 2023.\hskip 1em plus 0.5em minus 0.4em\relax New York, NY, USA: Association for Computing Machinery, 2023, p. 14–27.

\bibitem{stone2010opencl}
J.~E. Stone, D.~Gohara, and G.~Shi, ``Opencl: A parallel programming standard for heterogeneous computing systems,'' \emph{Computing in Science and Engg.}, vol.~12, no.~3, p. 66–73, may 2010.

\bibitem{reyes2020sycl}
R.~Reyes, G.~Brown, R.~Burns, and M.~Wong, ``Sycl 2020: More than meets the eye,'' in \emph{Proceedings of the International Workshop on OpenCL}, ser. IWOCL '20.\hskip 1em plus 0.5em minus 0.4em\relax New York, NY, USA: Association for Computing Machinery, 2020.

\bibitem{haris2023secda-tflite}
J.~Haris, P.~Gibson, J.~Cano, N.~{Bohm Agostini}, and D.~Kaeli, ``{SECDA-TFLite: A toolkit for efficient development of FPGA-based DNN accelerators for edge inference},'' \emph{Journal of Parallel and Distributed Computing}, vol. 173, pp. 140--151, 2023.

\bibitem{jiao2019tinybert}
X.~Jiao, Y.~Yin, L.~Shang, X.~Jiang, X.~Chen, L.~Li, F.~Wang, and Q.~Liu, ``Tinybert: Distilling bert for natural language understanding,'' \emph{arXiv preprint arXiv:1909.10351}, vol.~0, no.~0, pp. 1--12, 2019.

\bibitem{albericio2016cnvlutin}
J.~Albericio, P.~Judd, T.~Hetherington, T.~Aamodt, N.~E. Jerger, and A.~Moshovos, ``Cnvlutin: Ineffectual-neuron-free deep neural network computing,'' \emph{ACM SIGARCH Computer Architecture News}, vol.~44, no.~3, pp. 1--13, 2016.

\bibitem{zhang2016cambricon}
S.~Zhang, Z.~Du, L.~Zhang, H.~Lan, S.~Liu, L.~Li, Q.~Guo, T.~Chen, and Y.~Chen, ``Cambricon-x: An accelerator for sparse neural networks,'' in \emph{2016 49th Annual IEEE/ACM International Symposium on Microarchitecture (MICRO)}, IEEE.\hskip 1em plus 0.5em minus 0.4em\relax Taipei, Taiwan: IEEE, 2016, pp. 1--12.

\bibitem{chen2014dadiannao}
Y.~Chen, T.~Luo, S.~Liu, S.~Zhang, L.~He, J.~Wang, L.~Li, T.~Chen, Z.~Xu, N.~Sun \emph{et~al.}, ``Dadiannao: A machine-learning supercomputer,'' in \emph{2014 47th Annual IEEE/ACM International Symposium on Microarchitecture}, IEEE.\hskip 1em plus 0.5em minus 0.4em\relax Cambridge, UK: IEEE, 2014, pp. 609--622.

\bibitem{zhang2015optimizing}
C.~Zhang, P.~Li, G.~Sun, Y.~Guan, B.~Xiao, and J.~Cong, ``Optimizing fpga-based accelerator design for deep convolutional neural networks,'' in \emph{Proceedings of the 2015 ACM/SIGDA international symposium on field-programmable gate arrays}.\hskip 1em plus 0.5em minus 0.4em\relax New York, NY, USA: Association for Computing Machinery, 2015, pp. 161--170.

\bibitem{perfteamNDperf}
\BIBentryALTinterwordspacing
{The Linux Perf Team}, ``Perf wiki,'' n.d., accessed on April 13, 2023. [Online]. Available: \url{https://perf.wiki.kernel.org/index.php/Main_Page}
\BIBentrySTDinterwordspacing

\bibitem{arm2023neonregisters}
\BIBentryALTinterwordspacing
A.~Developer, ``Neon registers,'' 2023. [Online]. Available: \url{https://developer.arm.com/documentation/dht0002/a/Introducing-NEON/NEON-architecture-overview/NEON-registers}
\BIBentrySTDinterwordspacing

\bibitem{wang2017resnet}
F.~Wang, M.~Jiang, C.~Qian, S.~Yang, C.~Li, H.~Zhang, X.~Wang, and X.~Tang, ``Residual attention network for image classification,'' in \emph{Proceedings of the IEEE conference on computer vision and pattern recognition}, Honolulu, HI, USA, 2017, pp. 3156--3164.

\bibitem{wolf2020transformers}
\BIBentryALTinterwordspacing
T.~Wolf, L.~Debut, V.~Sanh, J.~Chaumond, C.~Delangue, A.~Moi, P.~Cistac, T.~Rault, R.~Louf, M.~Funtowicz, J.~Davison, S.~Shleifer, P.~von Platen, C.~Ma, Y.~Jernite, J.~Plu, C.~Xu, T.~Le~Scao, S.~Gugger, M.~Drame, Q.~Lhoest, and A.~Rush, ``Transformers: State-of-the-art natural language processing,'' in \emph{Proceedings of the 2020 Conference on Empirical Methods in Natural Language Processing: System Demonstrations}.\hskip 1em plus 0.5em minus 0.4em\relax Online: Association for Computational Linguistics, Oct. 2020, pp. 38--45. [Online]. Available: \url{https://aclanthology.org/2020.emnlp-demos.6}
\BIBentrySTDinterwordspacing

\bibitem{torchmlir2021torchmlir}
\BIBentryALTinterwordspacing
{Torch-MLIR Developers}, ``{The Torch-MLIR Project},'' 2021. [Online]. Available: \url{https://github.com/llvm/torch-mlir}
\BIBentrySTDinterwordspacing

\bibitem{Sozzo2018Frost}
E.~{Del Sozzo}, R.~{Baghdadi}, S.~{Amarasinghe}, and M.~D. {Santambrogio}, ``{A Unified Backend for Targeting FPGAs from DSLs},'' in \emph{ASAP}.\hskip 1em plus 0.5em minus 0.4em\relax Milan, Italy: IEEE, 2018, pp. 1--8.

\bibitem{agostini2022SODA}
N.~Bohm~Agostini, S.~Curzel, J.~Zhang, A.~Limaye, C.~Tan, V.~Amatya, M.~Minutoli, V.~G. Castellana, J.~Manzano, D.~Brooks, G.-Y. Wei, and A.~Tumeo, ``Bridging python to silicon: The soda toolchain,'' \emph{IEEE Micro}, vol.~42, no.~5, 2022.

\bibitem{agostini2022sodaopt}
N.~Bohm~Agostini, S.~Curzel, V.~Amatya, C.~Tan, M.~Minutoli, V.~G. Castellana, J.~Manzano, D.~Kaeli, and A.~Tumeo, ``An mlir-based compiler flow for system-level design and hardware acceleration,'' in \emph{ICCAD}.\hskip 1em plus 0.5em minus 0.4em\relax New York, NY, USA: Association for Computing Machinery, 2022.

\bibitem{stjerngren2022Bifrost}
A.~Stjerngren, P.~Gibson, and J.~Cano, ``Bifrost: {{End-to-End Evaluation}} and optimization of {{Reconfigurable DNN Accelerators}},'' in \emph{2022 {{IEEE International Symposium}} on {{Performance Analysis}} of {{Systems}} and {{Software}} ({{ISPASS}})}.\hskip 1em plus 0.5em minus 0.4em\relax Singapore: IEEE, May 2022, pp. 288--299.

\bibitem{xiang2022heteroflow}
S.~Xiang, Y.-H. Lai, Y.~Zhou, H.~Chen, N.~Zhang, D.~Pal, and Z.~Zhang, ``Heteroflow: An accelerator programming model with decoupled data placement for software-defined fpgas,'' in \emph{Proceedings of the 2022 ACM/SIGDA International Symposium on Field-Programmable Gate Arrays}, ser. FPGA '22.\hskip 1em plus 0.5em minus 0.4em\relax New York, NY, USA: Association for Computing Machinery, 2022, p. 78–88.

\bibitem{lai2019heterocl}
Y.-H. Lai, Y.~Chi, Y.~Hu, J.~Wang, C.~H. Yu, Y.~Zhou, J.~Cong, and Z.~Zhang, ``Heterocl: A multi-paradigm programming infrastructure for software-defined reconfigurable computing,'' in \emph{Proceedings of the 2019 ACM/SIGDA International Symposium on Field-Programmable Gate Arrays}, ser. FPGA '19.\hskip 1em plus 0.5em minus 0.4em\relax New York, NY, USA: Association for Computing Machinery, 2019, p. 242–251.

\bibitem{yang2020interstellar}
X.~Yang, M.~Gao, Q.~Liu, J.~Setter, J.~Pu, A.~Nayak, S.~Bell, K.~Cao, H.~Ha, P.~Raina, C.~Kozyrakis, and M.~Horowitz, ``Interstellar: Using halide's scheduling language to analyze dnn accelerators,'' in \emph{Proceedings of the Twenty-Fifth International Conference on Architectural Support for Programming Languages and Operating Systems}, ser. ASPLOS '20.\hskip 1em plus 0.5em minus 0.4em\relax New York, NY, USA: Association for Computing Machinery, 2020, p. 369–383.

\bibitem{dave2019dmazerunner}
S.~Dave, Y.~Kim, S.~Avancha, K.~Lee, and A.~Shrivastava, ``Dmazerunner: Executing perfectly nested loops on dataflow accelerators,'' \emph{ACM Trans. Embed. Comput. Syst.}, vol.~18, no.~5s, oct 2019.

\bibitem{cong2018polysa}
J.~Cong and J.~Wang, ``Polysa: Polyhedral-based systolic array auto-compilation,'' in \emph{2018 IEEE/ACM International Conference on Computer-Aided Design (ICCAD)}.\hskip 1em plus 0.5em minus 0.4em\relax San Diego, CA, USA: IEEE, 2018, pp. 1--8.

\bibitem{huang2021cosa}
Q.~Huang, M.~Kang, G.~Dinh, T.~Norell, A.~Kalaiah, J.~Demmel, J.~Wawrzynek, and Y.~S. Shao, ``Cosa: Scheduling by constrained optimization for spatial accelerators,'' in \emph{2021 ACM/IEEE 48th Annual International Symposium on Computer Architecture (ISCA)}.\hskip 1em plus 0.5em minus 0.4em\relax Valencia, Spain: IEEE, 2021, pp. 554--566.

\bibitem{vaidya2022sched}
M.~Vaidya, A.~Sukumaran-Rajam, A.~Rountev, and P.~Sadayappan, ``Comprehensive accelerator-dataflow co-design optimization for convolutional neural networks,'' in \emph{2022 IEEE/ACM International Symposium on Code Generation and Optimization (CGO)}.\hskip 1em plus 0.5em minus 0.4em\relax Seoul, South Korea: Association for Computing Machinery, 2022, pp. 325--335.

\bibitem{kao2020confuciux}
S.-C. Kao, G.~Jeong, and T.~Krishna, ``Confuciux: Autonomous hardware resource assignment for dnn accelerators using reinforcement learning,'' in \emph{2020 53rd Annual IEEE/ACM International Symposium on Microarchitecture (MICRO)}.\hskip 1em plus 0.5em minus 0.4em\relax Athens, Greece: IEEE, 2020, pp. 622--636.

\bibitem{zheng2020flextensor}
S.~Zheng, Y.~Liang, S.~Wang, R.~Chen, and K.~Sheng, ``Flextensor: An automatic schedule exploration and optimization framework for tensor computation on heterogeneous system,'' in \emph{Proceedings of the Twenty-Fifth International Conference on Architectural Support for Programming Languages and Operating Systems}, ser. ASPLOS '20.\hskip 1em plus 0.5em minus 0.4em\relax New York, NY, USA: Association for Computing Machinery, 2020, p. 859–873.

\bibitem{chen2018autotvm}
\BIBentryALTinterwordspacing
T.~Chen, L.~Zheng, E.~Yan, Z.~Jiang, T.~Moreau, L.~Ceze, C.~Guestrin, and A.~Krishnamurthy, ``Learning to optimize tensor programs,'' in \emph{Advances in Neural Information Processing Systems}, S.~Bengio, H.~Wallach, H.~Larochelle, K.~Grauman, N.~Cesa-Bianchi, and R.~Garnett, Eds., vol.~31.\hskip 1em plus 0.5em minus 0.4em\relax Curran Associates, Inc., 2018. [Online]. Available: \url{https://proceedings.neurips.cc/paper_files/paper/2018/file/8b5700012be65c9da25f49408d959ca0-Paper.pdf}
\BIBentrySTDinterwordspacing

\bibitem{zheng2020ansor}
L.~Zheng, C.~Jia, M.~Sun, Z.~Wu, C.~H. Yu, A.~Haj-Ali, Y.~Wang, J.~Yang, D.~Zhuo, K.~Sen \emph{et~al.}, ``{Ansor: Generating High-Performance tensor programs for deep learning},'' in \emph{14th USENIX symposium on operating systems design and implementation (OSDI 20)}, 2020, pp. 863--879.

\bibitem{min2023flexer}
H.~Min, J.~Kwon, and B.~Egger, ``Flexer: Out-of-order scheduling for multi-npus,'' in \emph{Proceedings of the 21st ACM/IEEE International Symposium on Code Generation and Optimization}, ser. CGO 2023.\hskip 1em plus 0.5em minus 0.4em\relax New York, NY, USA: Association for Computing Machinery, 2023, p. 212–223.

\bibitem{mlir2022transform}
\BIBentryALTinterwordspacing
M.~Developers, ``{Transform Dialect: Fine-grain transformation control dialect},'' 2022, online accessed on 11-04-2023. [Online]. Available: \url{https://mlir.llvm.org/docs/Dialects/Transform/}
\BIBentrySTDinterwordspacing

\end{thebibliography}

\end{document}